\documentclass[]{raa}           
\usepackage{graphicx,times}
\usepackage{natbib}
\usepackage{amssymb,amsmath}
\bibpunct{(}{)}{;}{a}{}{,}

\usepackage[a4paper=true,driverfallback=dvipdfm,pagebackref=true]{hyperref}
\hypersetup{pdftitle = The title of my PDF, pdfauthor = My name, pdfsubject= The subject, pdfkeywords = keyword1 keyword2 keyword3} 
\hypersetup{colorlinks = true, linkcolor = green, anchorcolor = red, citecolor = blue, filecolor = red, pagecolor = red, urlcolor = red}

\begin{document}

   \title{East Asian VLBI Network Observations of Active Galactic Nuclei Jets: Imaging with KaVA+Tianma+Nanshan
}

   \volnopage{Vol.0 (20xx) No.0, 000--000}      
   \setcounter{page}{1}          

   \author{Yu-Zhu Cui\inst{1,2}
   \and
    Kazuhiro Hada\inst{1,2}
    \and
    Motoki Kino\inst{2,3}
    \and
Bong-Won Sohn\inst{4,5,6}
\and
Jongho Park\inst{7}
\and
Hyun-Wook Ro\inst{4,5}
\and
Satoko Sawada-Satoh\inst{8}
\and
Wu Jiang\inst{9,11}
\and
Lang Cui\inst{10,11}
\and
Mareki Honma\inst{1,2,12}
\and
Zhi-Qiang Shen\inst{9,11}
\and
Fumie Tazaki\inst{2}
\and
Tao An\inst{9,11}
\and
Ilje Cho\inst{4,6}
\and
Guang-Yao Zhao\inst{4,13}
\and
Xiao-Peng Cheng\inst{4}
\and
Kotaro Niinuma\inst{8}
\and
Kiyoaki Wajima\inst{4}
\and
Ying-Kang Zhang\inst{9}
\and
Noriyuki Kawaguchi\inst{2}
\and
Juan-Carlos Algaba\inst{14}
\and
Shoko Koyama\inst{7}
\and
Tomoya Hirota\inst{2}
\and
Yoshinori Yonekura\inst{15}
\and
Nobuyuki Sakai\inst{4}
\and
Bo Xia\inst{9}
\and
Yong-Bin Jiang\inst{9}
\and
Lin-Feng Yu\inst{9}
\and
Wei Gou\inst{9}
\and
Ju-Yeon Hwang\inst{4}
\and
Yong-Chen Jiang\inst{9}
\and
Yun-Xia Sun\inst{9}
\and
Dong-Kyu Jung\inst{4}
\and
Hyo-Ryoung Kim\inst{4}
\and
Jeong-Sook Kim\inst{4}
\and
Hideyuki Kobayashi\inst{2}
\and
Jee-Won Lee\inst{4}
\and
Jeong-Ae Lee\inst{16}
\and
Hua Zhang\inst{10}
\and
Guang-Hui Li\inst{10}
\and
Zhi-Qiang Xu\inst{9}
\and
Peng Li\inst{10}
\and
Jung-Hwan Oh\inst{4}
\and
Se-Jin Oh\inst{4}
\and
Chung-Sik Oh\inst{4}
\and
Tomoaki Oyama\inst{2}
\and
Duk-Gyoo Roh\inst{4}
\and
Katsunori-M. Shibata\inst{2}
\and
Wen Guo\inst{9}
\and
Rong-Bing Zhao\inst{9}
\and
Wei-Ye Zhong\inst{9}
\and
Jin-Qing Wang\inst{9}
\and
Wen-Jun Yang\inst{10}
\and
Hao Yan\inst{10}
\and
Jae-Hwan Yeom\inst{4}
\and
Bin Li\inst{9}
\and
Xiao-Fei Li\inst{10}
\and
Jian-Ping Yuan\inst{10}
\and
Jian Dong\inst{9}
\and
Zhong Chen\inst{9}
\and
Kazunori Akiyama\inst{17,18}
\and
Keiichi Asada\inst{7}
\and
Do-Young Byun\inst{4}
\and
Yoshiaki Hagiwara\inst{19}
\and
Jeffrey Hodgson\inst{4}
\and
Tae-Hyun Jung\inst{4}
\and
Kee-Tae Kim\inst{4}
\and
Sang-Sung Lee\inst{4}
\and
Kunwoo Yi\inst{20}
\and
Qing-Hui Liu\inst{9}
\and
Xiang Liu\inst{10,11}
\and
Ru-Sen Lu\inst{9,11,21}
\and
Masanori Nakamura\inst{7}
\and
Sascha Trippe\inst{20}
\and
Na Wang\inst{10,11}
\and
Xue-Zheng Wang\inst{9}
\and
Bo Zhang\inst{9}
   }

   \institute{Department of Astronomical Science, The Graduate University for Advanced Studies, SOKENDAI, 2-21-1 Osawa, Mitaka, Tokyo 181-8588, Japan; {\it yuzhu.cui@grad.nao.ac.jp}\\
        \and
        Mizusawa VLBI Observatory, National Astronomical Observatory of Japan, 2-12 Hoshigaoka, Mizusawa, Oshu, Iwate 023-0861, Japan\\   
        \and
        Kogakuin University of Technology and Engineering, Academic Support Center, 2665-1 Nakano, Hachioji, Tokyo 192-0015, Japan\\   
        \and
        Korea Astronomy and Space Science Institute, Yuseong-gu, Daejeon 34055, Korea\\   
        \and
        Department of Astronomy, Yonsei University, 50 Yonsei-ro, Seodaemun-gu, Seoul 03722, Korea\\   
        \and
        Department of Astronomy and Space Science, University of Science and Technology, 217 Gajeong-ro, Daejeon, Korea\\
        \and
        Institute of Astronomy and Astrophysics, Academia Sinica, 645 N Aohoku Pl, Hilo, HI 96720, USA\\   
        \and
        Graduate School of Sciences and Technology for Innovation, Yamaguchi University, Yoshida 1677-1, Yamaguchi, Yamaguchi 753-8512, Japan\\   
        \and
        Shanghai Astronomical Observatory, Chinese Academy of Sciences, 80 Nandan Road, Xuhui District, Shanghai 200030, China\\   
        \and
        Xinjiang Astronomical Observatory, Chinese Academy of Sciences, Urumqi 830011, China\\   
        \and
        Key Laboratory of Radio Astronomy, Chinese Academy of Sciences, Nanjing 210008, China\\   
        \and
        Institute of Astronomy, The University of Tokyo, 2-21-1 Osawa, Mitaka, Tokyo 181-0015, Japan\\   
        \and
        Instituto de Astrofísica de Andalucía-CSIC, Glorieta de la Astronomía s/n, Granada E-18008, Spain\\   
        \and
        Department of Physics, Faculty of Science, University of Malaya, Kuala Lumpur 50603, Malaysia\\   
        \and
        Center for Astronomy, Ibaraki University, 2-1-1 Bunkyo, Mito, Ibaraki 310-8512, Japan\\   
        \and
        Space Light Laboratory, 77, Yangjaedae-ro 86 road, Gangdong-gu, Seoul, Korea\\   
        \and
        Massachusetts Institute of Technology Haystack Observatory, 99 Millstone Road, Westford, MA 01886, USA\\  
        \and
        National Radio Astronomy Observatory, 520 Edgemont Rd, Charlottesville, VA 22903, USA\\   
        \and
        Toyo University, 5-28-20 Hakusan, Bunkyo-ku, Tokyo 112-8606, Japan\\ 
        \and
        Department of Physics and Astronomy, Seoul National University, Gwanak-gu, Seoul 08826, Korea\\
        \and
        Max Planck Institute for Radio Astronomy, Auf dem h\"ugel 69, Bonn 53121, Germany\\
\vs\no
   {\small Received~~2021 March 29; accepted~~2021~~April 11}}

\abstract{ The East Asian very-long-baseline interferometry (VLBI) Network (EAVN) is a rapidly evolving international VLBI array that is currently promoted under joint efforts among China, Japan, and Korea. EAVN aims at forming a joint VLBI Network by combining a large number of radio telescopes distributed over East Asian regions. After the combination of the Korean VLBI Network (KVN) and the VLBI Exploration of Radio Astrometry (VERA) into KaVA, further expansion with the joint array in East Asia is actively promoted. Here we report the first imaging results (at 22 and 43\,GHz) of bright radio sources obtained with KaVA connected to Tianma 65-m and Nanshan 26-m Radio Telescopes in China. To test the EAVN imaging performance for different sources, we observed four active galactic nuclei (AGN) having different brightness and morphology. As a result, we confirmed that Tianma 65-m Radio Telescope (TMRT) significantly enhances the overall array sensitivity, a factor of 4 improvement in baseline sensitivity and 2 in image dynamic range compared to the case of KaVA only. The addition of Nanshan 26-m Radio Telescope (NSRT) further doubled the east-west angular resolution. With the resulting high-dynamic-range, high-resolution images with EAVN (KaVA+TMRT+NSRT), various fine-scale structures in our targets, such as the counter-jet in M\,87, a kink-like morphology of the 3C\,273 jet and the weak emission in other sources, are successfully detected. This demonstrates the powerful capability of EAVN to study AGN jets and to achieve other science goals in general. Ongoing expansion of EAVN will further enhance the angular resolution, detection sensitivity and frequency coverage of the network.  
}

   \authorrunning{Yuzhu Cui et al. }            
   \titlerunning{EAVN Observations of Active Galactic Nuclei Jets}  

   \maketitle

%
%
\section{Introduction}
Very Long Baseline Interferometry (VLBI) is a powerful astronomical technique to observe radio sources at a high angular resolution. The recent successful detection of the black-hole shadow in M\,87 with the Event Horizon Telescope (EHT) is an excellent example of such a technique at millimeter wavelengths (\citealt{eht2019a}). VLBI also plays a unique role in resolving the formation scales of relativistic jets powered by active galactic nuclei (AGN) (e.g., \citealt{hada2020}). Moreover, high-resolution VLBI can also precisely determine the spatial distributions, kinematics and parallaxes of astrophysical masers, which provides insights into the structures of star forming regions and our Galaxy (e.g., \citealt{reid2014, hirota2020}). To further promote such VLBI sciences, it is essential to enhance the angular resolution, the sensitivity and the imaging performance of the network. This requires the establishment of a large VLBI array across multiple countries. 

\begin{figure}[htbp]
 \begin{center}
  \includegraphics[width=0.8\textwidth]{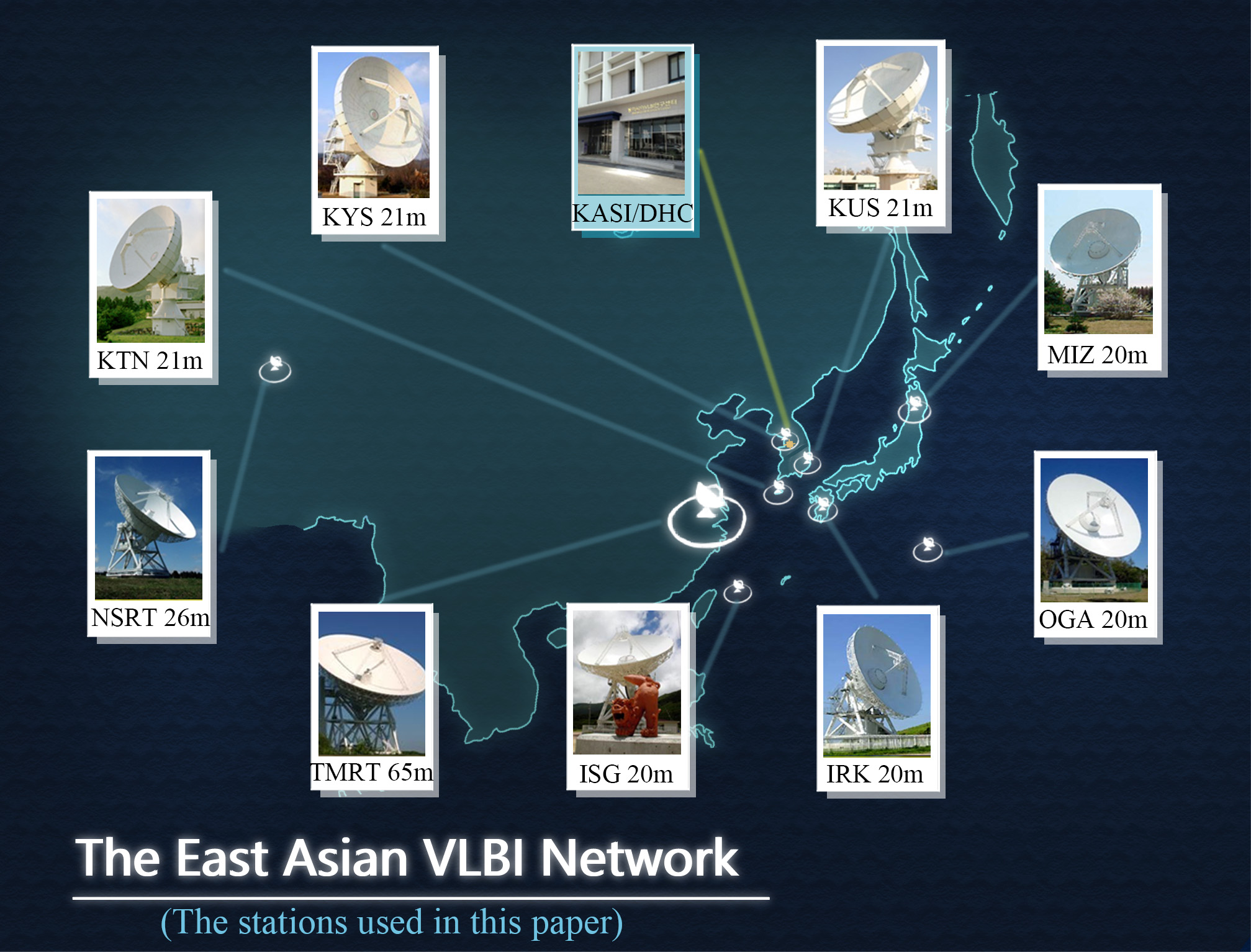}
 \end{center}
 \caption{Geographic distribution map of the radio telescopes used in this paper. In addition, the correlation center in KASI/Daejeon is also shown. The size of the cartoon antenna on the map is proportional to the diameter of the dish.}\label{fig:EAVN}
\end{figure}

In recent years, international collaboration of VLBI facilities in East Asia is rapidly growing. In particular, the Korean VLBI Network (KVN; e.g., \citealt{lee2014}) in Korea and the VLBI Exploration of Radio Astrometry (VERA; e.g., \citealt{kobayashi2003}) in Japan were successfully combined into a single network at 22 and 43\,GHz as the KVN and VERA Array (KaVA), and now the joint array has been in regular operation since 2014 (e.g., \citealt{sawada2013}). KaVA significantly improved the imaging performance for bright AGN jets compared to the ones obtained with the individual arrays thanks to the increased numbers of stations and baselines (e.g., \citealt{niinuma2014}). Since then a growing number of publications based on KaVA have been produced for a variety of science targets (e.g., \citealt{matsumoto2014, dodson2014, oh2015, kim2015, yun2016, an2016, hada2017, zhang2017, burns2018, zhao2019, lee2019, imai2019, baek2019, park2019, hada2020}). 

Nevertheless, the number of baselines, angular resolution and resulting fringe/image sensitivity of KaVA are still limited. To further expand the international VLBI array in East Asia, the addition of stations from the Chinese VLBI Network (CVN; (e.g., \citealt{ye1991}) is certainly important. CVN is operated under the auspices of the Chinese Academy of Sciences (CAS), and currently, it consists of 4 primary stations: Tianma 65-m Radio Telescope in Shanghai (hereafter, TMRT), Nanshan 26-m Radio Telescope in Urumqi (hereafter, NSRT), Miyun 50-m radio telescope in Beijing and Kunming 40-m Telescope in Yunnan. The large collecting area and wide geographically distributed stations over the mainland in China are very complementary to VLBI facilities in Korea and Japan. Therefore, combining KaVA and CVN (as well as other radio telescopes in these countries such as the Japanese VLBI Network (JVN; e.g., \citealt{doi2006, yonekura2016}), Sejong station in Korea (e.g., \citealt{kondo2008}) and Five-hundred-meter Aperture Spherical radio Telescope in China (FAST; e.g., \citealt{nan2006}) can greatly enhance the angular resolution and overall sensitivity of the network. 

\begin{table}
\begin{center}
\caption[]{General information of the nine stations from EAVN array used in this paper. The parameters listed here are from the EAVN status report. $^{a}$ Diameter in meter. $^{b}$ Latitude in degree. $^{c}$ Longitude in degree. $^{d}$ Altitude in meter. $^{e}$ Aperture efficiency. $^{f}$ HPBW is Half Power Beam Width in arcsecond.}\label{tab:stations}
 \begin{tabular}{lcccccccccc}
  \hline\noalign{\smallskip}
    Location  & Name&Network & D$^{a}$ & Lat.$^{b}$& Lon.$^{c}$& Alt.$^{d}$  &  \multicolumn{2}{c}{$\eta^{e}$ (\%)} & \multicolumn{2}{c}{HPBW$^{f}$}  \\
            &   &       &  (m)& (degree) & (degree)   &  (m)&22\,GHz& 43\,GHz&22\,GHz& 43\,GHz\\
    \noalign{\smallskip}\hline
    Tianma      &TMRT   & CVN  & 65 &  31  & 121  & 49  & 50  &  45  & 44  & 22         \\
    Nanshan       &NSRT & CVN  & 26 &  43  & 87  & 2029 & 60  &  -   & 115 & -      \\
    \noalign{\smallskip}\hline
    Mizusawa    &MIZ   & VERA & 20 &  39  & 141  & 116  & 47  & 51   & 141 & 71        \\
    Iriki        &IRK  & VERA & 20 &  32  & 130  & 574  & 47  & 44   & 149 & 78        \\
    Ogasawara    &OGA  & VERA & 20 &  272  & 142  & 273  & 50  & 45   & 143 & 78        \\
    Ishigakijima  &ISG & VERA & 20 &  242  & 124  & 65   & 49  & 48   & 144 & 79          \\
    Tamna         &KTN & KVN  & 21 &  332 & 126  & 452  & 60  & 63   & 126 & 63        \\
    Ulsan         &KUS & KVN  & 21 &  36  & 129  & 170  & 63  & 61   & 124 & 63         \\
    Yonsei       &KYS  & KVN  & 21 &  38  & 127  & 139  & 55  & 63   & 127 & 63        \\
  \noalign{\smallskip}\hline
\end{tabular}
\end{center}
\end{table}

The concept of the East Asian VLBI Network (EAVN) was first discussed in 2003 (e.g., \citealt{shen2004}) and the consortium to facilitate EAVN was established in 2004 (e.g., \citealt{inoue2005}). Since then, significant joint efforts have been made among China, Japan and Korea to promote EAVN activities, and some early EAVN experiments were already made in 2010 (e.g., \citealt{fujisawa2014, sugiyama2016, wajima2016}). However, in these early days, the observations were temporarily coordinated with ad-hoc arrays and also KaVA was still under commissioning. Observations with EAVN became more organized and intensive from 2016, when KaVA was already in stable operation, TMRT started its early science operation and NSRT was back in operation after its refurbishment (see section~\ref{sec:array}). To accelerate the commissioning of KaVA+TMRT+NSRT which serves as a core array of EAVN, we performed a large EAVN observing campaign between March and May 2017. The EAVN campaign was performed by making use of the slots allocated to the KaVA AGN Large Program that intensively monitored the nearby supermassive black holes M\,87 and Sgr\,A$^\star$ at 22 and 43\,GHz (\citealt{kino2015}). TMRT and NSRT regularly participated in these KaVA sessions (and occasionally, some more stations joined such as Hitachi, Takahagi, Kashima, Nobeyama, Sejong, Medicina and Noto), resulting in the largest EAVN experiments that have ever made. Therefore, these datasets are very useful to evaluate the array performance of EAVN as well as to study the physics close to the supermassive black holes (SMBHs).  

\begin{table}
\begin{center}
\caption[]{ Array Specifications. $^a$ Array symbols for different combination of stations. $^b$ Number of antennas. $^c$ Number of baselines. $^d$ Baseline length in km. $^e$ Angular resolution in milli-arcsecond. $^f$ Image sensitivity $\sigma_{\rm I}$ is the typical value adopted in the EAVN status report under the assumption of an integration time $t$ = 4\,hours and a total bandwidth $B$ = 256\,MHz. }\label{tab:ims}


 \begin{tabular}{llccccccccc}
  \hline\noalign{\smallskip}
    Array$^a$ &Stations & $N_{\rm Ant}^b$ & $N_{\rm bl}^c$ &\multicolumn{2}{c}{ $L_{\rm bl}^d$ (km)} & \multicolumn{2}{c}{ $\theta^e$ (mas)}  & \multicolumn{2}{c}{$\sigma_{\rm I}^f$ ($\mu$Jy/beam)}    \\
            && &  &min & max & 22\,GHz & 43\,GHz & 22\,GHz & 43\,GHz  \\
    \hline\noalign{\smallskip}
    A1&KaVA                      & 7  & 21 & 305& 2270      & 1.24  & 0.63  & 155 & 268\\
    A2&KaVA+TMRT           & 8  & 28 & 305& 2270      & 1.24  & 0.63  & 95  & 165\\
    A3&KaVA+TMRT+NSRT & 9  & 36 & 305& 5078      & 0.55  & --  &   75    & --\\
  \noalign{\smallskip}\hline
\end{tabular}
\end{center}
\end{table}

In this paper, we report the initial results of EAVN imaging observations for several AGN based on a subset of the EAVN-2017 campaign data. While here we focus on the performance evaluation of the KaVA+TMRT+NSRT array, more dedicated scientific results/analysis on individual sources using the whole campaign dataset will be reported in separate papers. In section~\ref{sec:array}, we overview the VLBI network used in this paper, with a special emphasis on TMRT and NSRT. The basic information about the observed sources is presented in section~\ref{sec:source}. Section~\ref{sec:obs} includes the detailed information of the observations and data reduction processes. The results and discussion are described in section~\ref{sec:result}. EAVN data status after 2017 is briefly summarized in section~\ref{sec:2021}. In the last section, we summarize the paper and the future plan of EAVN. Throughout this paper, we assume a flat $\Lambda$CDM universe with the cosmological parameters from \citet{planck2016}, $H_0$ = 67.8\,km\,s$^{-1}$\,Mpc$^{-1}$, $\Omega_{\rm m}$ = 0.31, and $\Omega_{\rm \Lambda}$ = 0.69.

\begin{table}
\begin{center}
\caption[]{ EAVN observations presented in this paper. Both of these two epoch used the observing mode as 32\,MHz$\times$ 8\,IFs. The main target is M\,87.}\label{tab:obs}


 \begin{tabular}{lcccccl}
  \hline\noalign{\smallskip}
     Obs. Code & Frequency (GHz) & Date  &   UT Time&  Stations   \\
    \hline
    a17077a (Session-K) &22  & 2017 March 18 & 12:45--19:45    &  KaVA (no KYS), TMRT, NSRT\\
    a17086a (Session-Q) &43  & 2017 March 27 & 13:10--18:10    & KaVA, TMRT\\
  \noalign{\smallskip}\hline
\end{tabular}
\end{center}
\end{table}

\section{EAVN array}
\label{sec:array}
The EAVN array is continuously expanding and the number of participating stations are increasing year by year. In the EAVN-2017 campaign, while a total of 15 stations joined from East Asia and Europe, the 9 stations of KaVA, TMRT and NSRT participated in the campaign on a regular basis. Figure~\ref{fig:EAVN} shows the geographical distribution of these sites and relative diameter. Table~\ref{tab:stations} adopted from the EAVN status report \footnote{https://radio.kasi.re.kr/eavn/files/Status$\_$Report$\_$EAVN$\_$2020A$\_$20191031.pdf} summarizes the basic information of the 9 stations which form a core array of EAVN. Below we describe some more details about TMRT and NSRT. For details of KaVA stations, see \citet{niinuma2014} and \citet{hada2017}.  

\begin{figure*}[ttt]
\begin{center}
\includegraphics[width=0.37\textwidth]{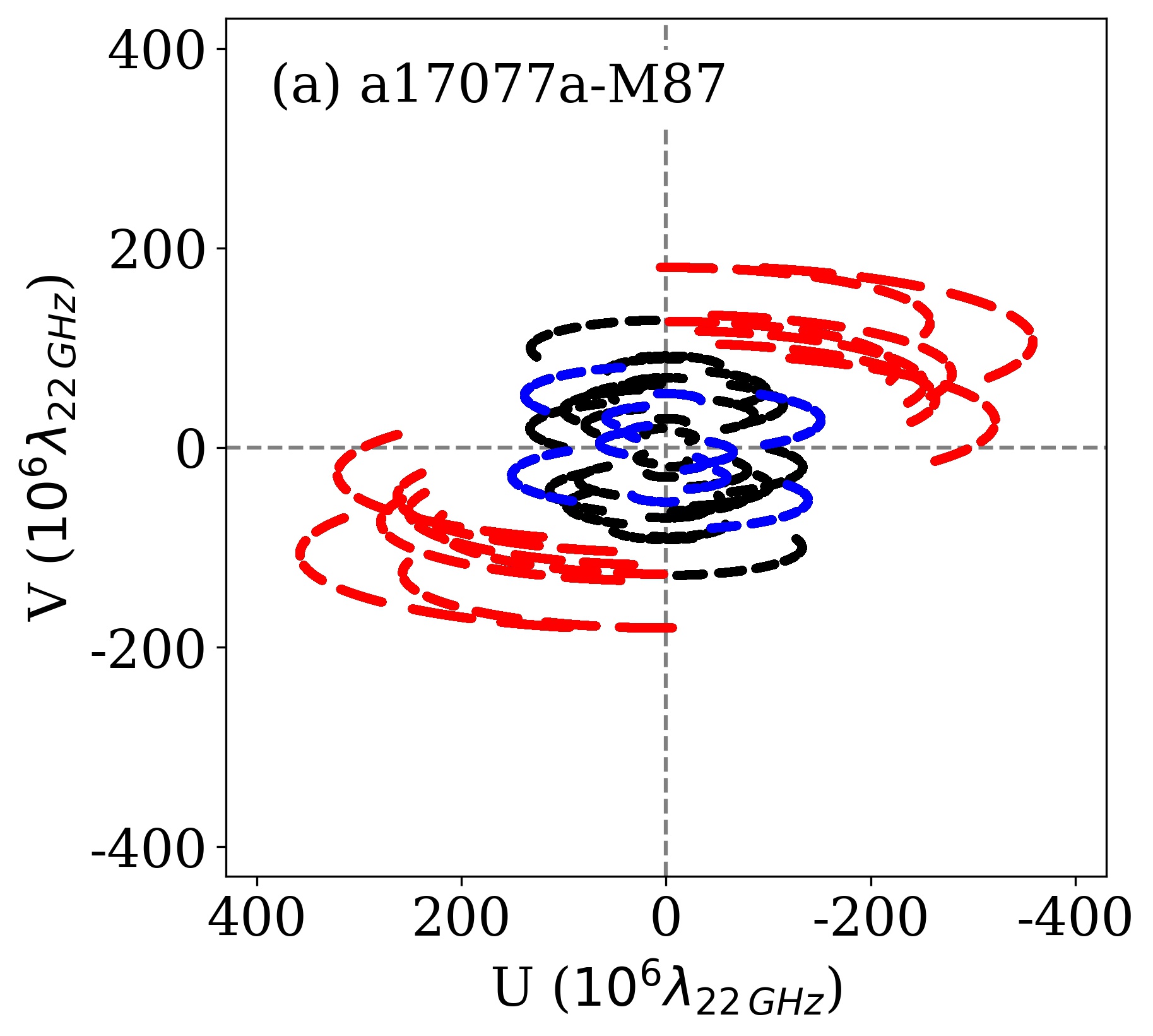}
\includegraphics[width=0.62\textwidth]{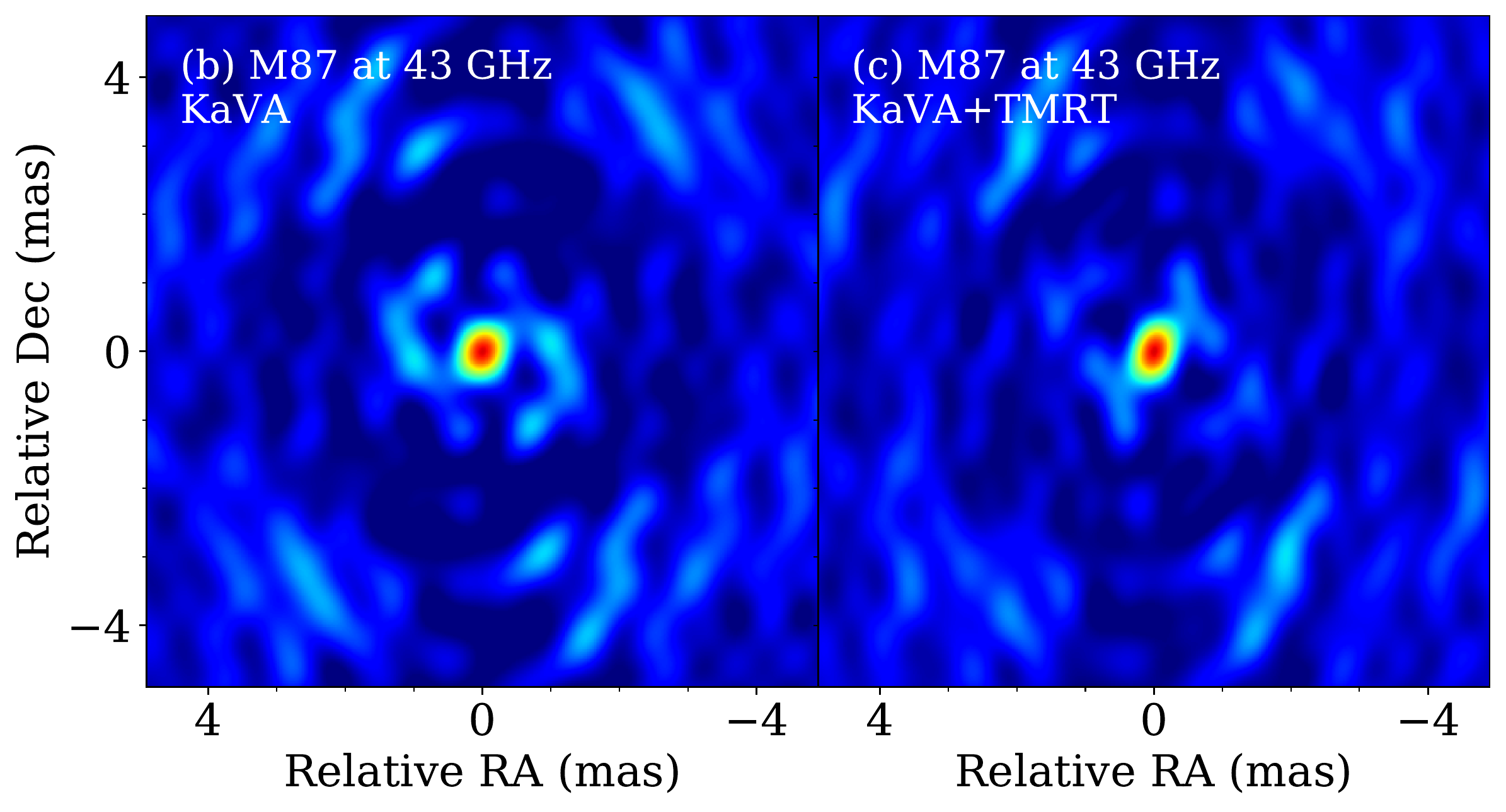}\\
\includegraphics[width=0.99\textwidth]{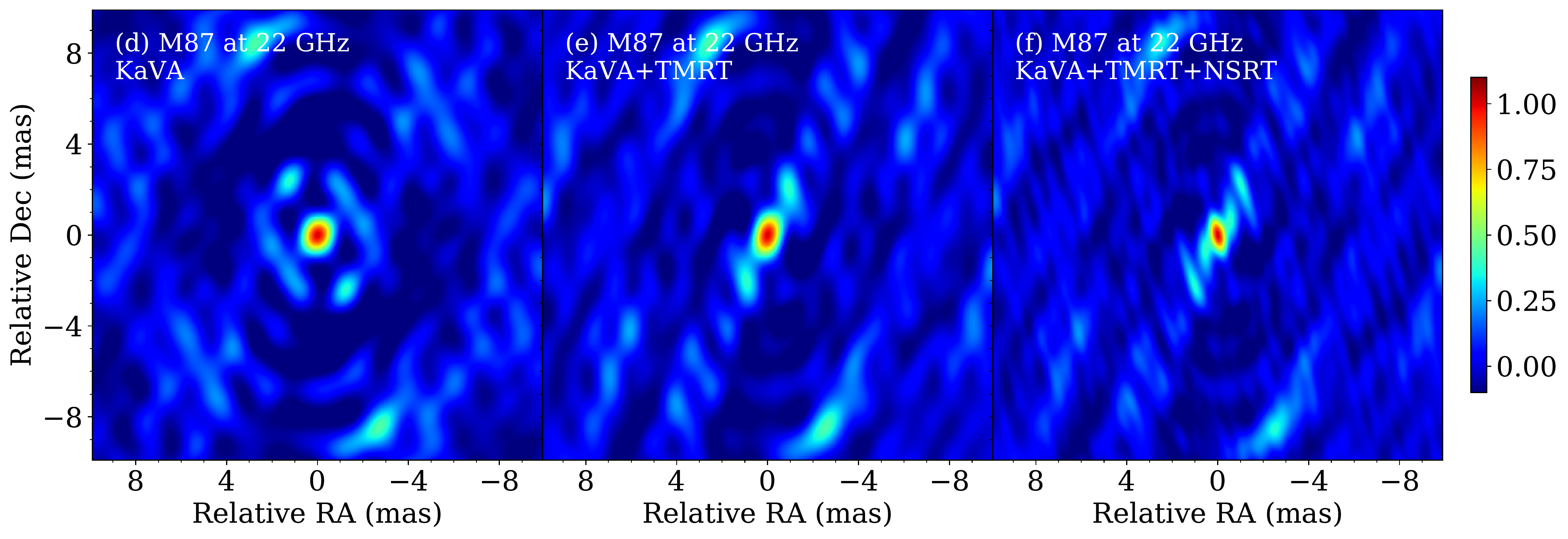} 
\end{center}
\caption{(a) The \emph{uv}-coverage for M\,87 in Session-K. Curves with blue, red and black color indicate baselines related to TMRT, NSRT and among only KaVA, respectively. (b)-(c): Dirty beam patterns with a naturally-weighted scheme for KaVA and KaVA+TMRT in Session-Q, respectively. (d)-(f): Dirty beam patterns with a naturally-weighted scheme for KaVA, KaVA+TMRT and KaVA+TMRT+NSRT in Session-K, respectively. }
\label{fig:uv}
\end{figure*}

\subsection{Tianma 65-m Radio Telescope (TMRT)}
The Tianma 65-m Radio Telescope (TMRT) is operated by the Shanghai Astronomical Observatory (SHAO) as a joint project between the Chinese Academy of Sciences (CAS) and the city of Shanghai. The project was approved at the end of October 2008. After laying the foundation in December 2009, the construction of the telescope started in March 2010 and completed with the first-light detection in October, 2012. The location of the telescope is in Sheshan, Songjiang District of Shanghai (\citealt{jiang2018}). 

Cooperated with another 25-m-diameter radio telescope in Sheshan, the first interferometric fringes of TMRT were detected with a 6.1-km baseline in November 2012. After that, the TMRT starts commissioning and makes great contributions to a broad range of radio astronomy researches such as blazars, microquasars, molecular spectral lines, pulsars, X-ray binaries and geodynamics based on both single-dish and VLBI modes (e.g. \citealt{li2016, yang2019, hou2020}). As a representative radio telescope in China, TMRT joins EAVN from the beginning of this project since 2013. TMRT also regularly joins European VLBI Network (EVN) since 2014 in particular at low frequencies (e.g. 5 GHz). 

At present, TMRT is the largest fully steerable telescope in East Asia. It has a wide range of observing wavelengths covering 1--50\,GHz with 8 bands, namely L (1.25--1.75\,GHz), S (2.2--2.4\,GHz), C (4--8\,GHz), X (8.2--9.0\,GHz), Ku (12--18\,GHz), K (18.0--26.5\,GHz), Ka (30--34\,GHz) and Q (35--50\,GHz) bands (\citealt{yan2015}). The S/X- and X/Ka-band receivers are the dual-frequency co-axis feed. The common observing frequencies of TMRT with KaVA are 22\,GHz and 43\,GHz (and partially also 6.7\,GHz). The maximum aperture efficiency of TMRT assembled to be reached at 53$^\circ$ elevation is around 50\% at 22\,GHz and 45\% at 43\,GHz with the active surface control system (see table~\ref{tab:stations}). The active surface controller is set `ON' by default. The nominal root mean square (rms) of surface accuracy is 0.6\,mm without active surface and 0.3\,mm with active surface at the elevation around 53$^\circ$. The sub-reflector surface accuracy is 0.1\,mm rms. The typical pointing accuracy is 3\,arcseconds when the wind speed is 4\,m\,s$^{-1}$ and 30\,arcseconds if the wind speed reaches 20\,m\,s$^{-1}$. The slew rate is 0.5\,$^\circ$\,s$^{-1}$ and 0.3\,$^\circ$\,s$^{-1}$ at azimuth and elevation directions respectively. Dual-pixel receivers are installed in TMRT at both 22 and 43\,GHz which enables simultaneous observations of multiple lines (\citealt{zhong2018}). These two beams have a fixed separation angle of 140\,arcseconds at 22\,GHz and 100\,arcseconds at 43\,GHz. One of the beams is placed at the antenna focus for VLBI observations. The Half Power Beam Width (HPBW) is the measured beam sizes listed in table~\ref{tab:stations}.

\subsection{Nanshan 26-m Radio Telescope (NSRT)}
The Nanshan 26-m Radio Telescope in Urumqi (NSRT) operated by Xinjiang Astronomical Observatory is another key station in CVN constructed in 1993. The telescope is located at the northern foot of Tianshan with high elevation above 2029\,m and over 75\,km away from Urumqi city which is surrounded by the superior observing environment with dry climate and low-level radio frequency interference. Originally the telescope was designed with a diameter of 25\,m, but a refurbishment of the telescope was made from early 2014 and completed in late 2015 (\citealt{xu2018}). This resulted in an enlargement of the main reflector to 26\,m and improvement of the antenna surface accuracy. After this reconstruction was finished, NSRT participated in EAVN commissioning. NSRT also joins EVN since early 1994 and International VLBI Service for Geodesy and Astrometry (IVS) after 1996. Over the past 20 years, NSRT has made rich contributions to pulsar timing and physics, molecular line survey and star formation, AGN jets and flux monitoring based on both single-dish and VLBI modes (e.g. \citealt{wang2005, yuan2010, wu2018, cui2010, liu2012}). 

NSRT features a Cassegrain-type design with a 26-m diameter main reflector and a 3-m sub-reflector on Az-El mount. Receivers at six frequency bands, L, S/X, C, K (22--24.2\,GHz), and Q (30--50\,GHz), are equipped, while the new cryogenic 43\,GHz receiver has been installed in 2018 and is under evaluation. The surface accuracy is 0.4\,mm (rms) for main-reflector and 0.1\,mm (rms) for sub-reflector. The slewing rates of the main reflector are $1.0\,^\circ$\,s$^{-1}$ in azimuth and $0.5\,^\circ$\,s$^{-1}$ in elevation. The pointing precision is 10\,arcseconds (rms). The aperture efficiency of NSRT is $60\%$ at 22\,GHz (see table~\ref{tab:stations}).

Thanks to its unique location at the northwest of China, the east-west baseline coverage of EAVN is significantly enhanced with the participation of NSRT (from 2270\,km to 5078\,km). This offers a fringe spacing (namely angular resolution $\theta = \lambda / D$, $\lambda$ is the observing wavelength and $D$ is the maximum baseline length) down to 0.55\,milliarcseconds (mas) at 22\,GHz (and 0.26\,mas at 43\,GHz), which is 2.3 times smaller than KaVA only (see table~\ref{tab:ims}).

\begin{figure}[htbp]
 \begin{center}
 \includegraphics[width=0.7\textwidth]{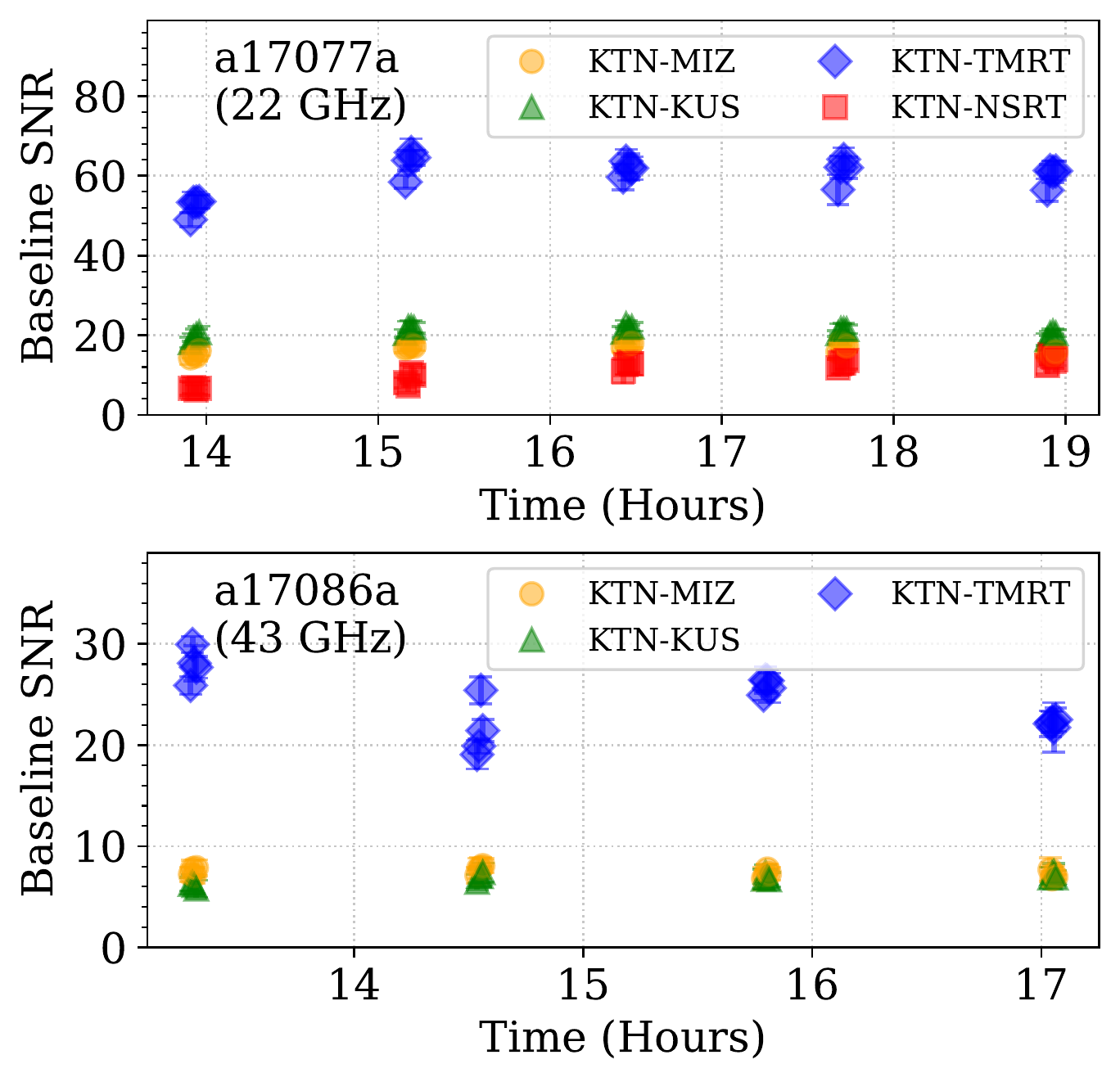}
 \end{center}
 \caption{Baseline signal-to-noise ratio ($SNR_{\rm {ij}}$) (output from {\tt FRING}) for various baselines for the point source 1219+044. (Upper panel) Session-K; (Lower panel) Session-Q. }
 \label{fig:snr}
\end{figure}

\section{Source selection}
\label{sec:source}
To evaluate the imaging performance of the KaVA+TMRT+NSRT array, we selected four well-known AGN sources, i.e., 1219+044, 3C\,273, M\,84 and M\,87. These sources were selected for the following reasons: 1) the core flux densities are high enough to detect fringes, although M\,84 is a possible exception; 2) the mas-scale radio morphology covers diverse structures from point-like to complicated; and 3) the sky positions are close with each other, which offers similar observing/calibration conditions (e.g., atmosphere, gain curves) among these sources. 

1219+044 (PKS\,J1222+0413) is a bright compact blazar located at a redshift of $z$ = 0.966 (1\,mas = 8.15\,pc, \citealt{paris2017}). The source has recently been identified as a candidate of $\gamma$-ray emitting narrow-line Seyfert I galaxy (\citealt{kynoch2019}). Previous VLBA images at 15\,GHz show an extremely compact core-dominated structure (unresolved up to 442$\times 10^6\lambda_{\mathrm 15GHz}=$ 8840\,km) with a tiny amount of jet emission toward the south (\citealt{lister2019}). Therefore 1219+044 may serve as a useful reference source to check the amplitude accuracy of each antenna.

3C\,273 ($z$ = 0.158, 1\,mas = 2.82\,pc, \citealt{strauss1992}) is one of the most famous quasars (\citealt{schmidt1963}) with a powerful relativistic jet. Due to strong relativistic boosting, on mas scales the source shows a one-sided jet towards the south-west direction at a position angle PA = $-138^\circ$ (\citealt{perley2017}). The viewing angle of the jet is estimated to be $\sim$3.8$^\circ$--7.2$^\circ$ (\citealt{meyer2016}). The mas-scale jet structure is rather complicated (knotty and helically twisted) with frequent ejections of new components from the core (\citealt{jorstad2017}). 

M\,84 ($z$ = 0.00339, 1\,mas = 0.07\,pc, \citealt{meyer2018}) is a nearby low-luminosity elliptical galaxy in the Virgo cluster. At radio, the source exhibits Fanaroff-Riley type I jet morphology on kpc scales. The quasi-symmetric morphology of the jet and counter-jet indicates a jet viewing angle to be close to edge-on (74$^{+9}_{-18}$)$^\circ$ (\citealt{meyer2018}). On mas scales, past VLBI images show relatively compact morphology with a slight extension towards the north (\citealt{giovannini2001, ly2004}). The radio core flux density is typically $\sim$100\,mJy or less. Therefore, M\,84 provides a useful reference to check EAVN imaging performance for low luminosity objects. 

\begin{figure}[htbp]
 \begin{center}
  \includegraphics[width=1\textwidth]{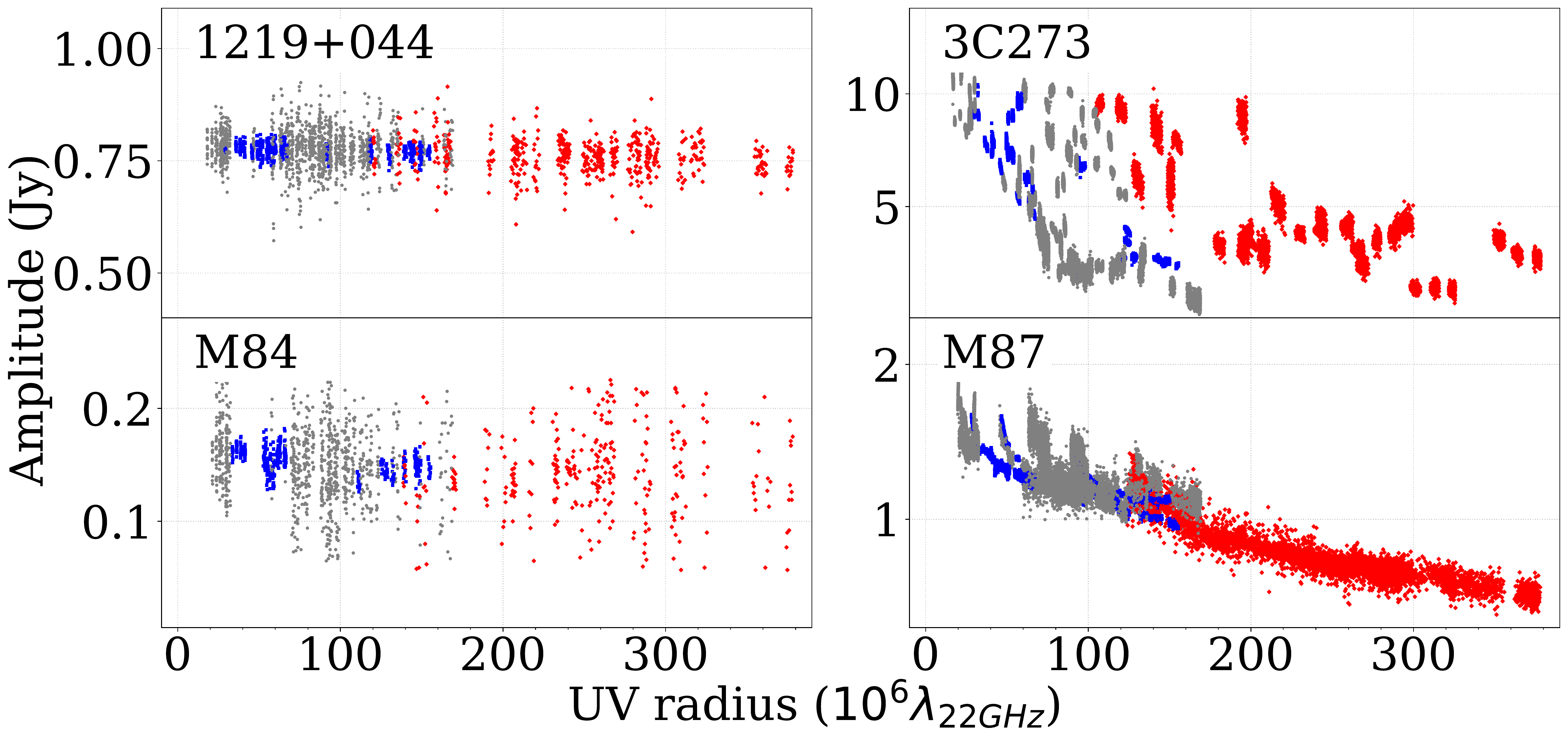}
    \end{center}
 \caption{Visibility amplitude distributions as a function of uv-distance for Session-K. Red, blue and gray points indicate the data related to NSRT, TMRT and KaVA only, respectively. The visibility data are averaged every 2\,mins.}
 \label{fig:amp17077K}
\end{figure}

M\,87 ($z$ = 0.00436, 1\,mas = 0.09\,pc, \citealt{smith2000}) is the giant radio galaxy at the center of the Virgo cluster with a prominent jet. The galaxy hosts a SMBH of $M_{\rm BH}$ $\sim$6.5 $\times 10^9 M_{\rm \odot}$ (\citealt{eht2019b}). On mas scales the source exhibits a highly collimated jet at PA $\sim$290$^\circ$ (\citealt{hada2016, walker2018}). In contrast to 3C\,273, the VLBI morphology of the M\,87 jet is relatively smooth without prominent knotty features. Since M\,87 is a primary target of the KaVA/EAVN Large Program (\citealt{kino2015}), a large fraction of the observing time was spent on this source in our observations.  

\begin{figure}[ttt]
 \begin{center}
  \includegraphics[width=0.85\textwidth]{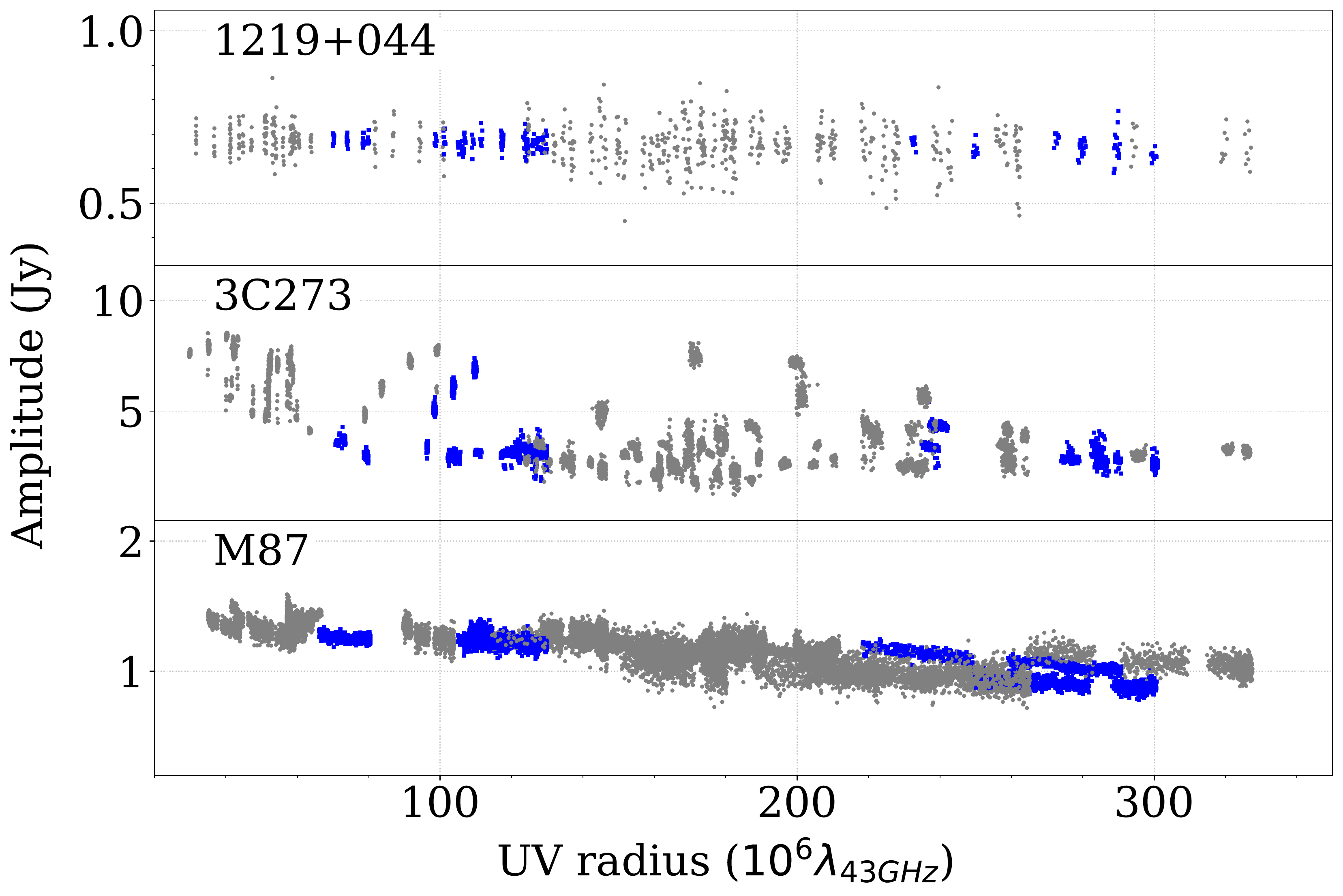}
    \end{center}
 \caption{Same as figure~\ref{fig:amp17077K} but for Session-Q. The visibility data are averaged every 1\,mins.}
 \label{fig:amp17086Q}
\end{figure}

\section{Observations and Data Reduction}
\label{sec:obs}
\subsection{Observations}
Here we selected two representative epochs from the EAVN 2017 campaign as summarized in table~\ref{tab:obs}. These observations were conducted on March 18th (observing code: a17077a; hereafter, Session-K) and March 27th (observing code: a17086a; hereafter, Session-Q) at 22 and 43\,GHz, respectively. Yonsei (KYS) did not join Session-K due to an issue at the site. TMRT participated in both epochs along with KaVA, while NSRT was only available at 22\,GHz. The overall weather condition was good at each site. For TMRT, antenna pointing calibration was made at the beginning of each session. We observed M\,87 as a primary target of the EAVN campaign while the other sources (3C\,273, 1219+044, M\,84) were observed with less on-source time. In detail, Session-K lasted for 7 hours where scans of 3C\,273 (6 min) -- 1219+044 (4 min) -- M\,87 (47 min) -- M\,84 (4 min) were repeated for 6 cycles. Session-Q were performed for 5 hours where scans of 3C\,273 (6 min) -- 1219+044 (2 min) -- M\,87 (54 min) -- M\,84 (2 min) --  were repeated for 4 cycles. RT Vir (H$_{2}$O/SiO masers) was inserted with four scans (4 min / each scan) for a system/frequency check of both experiments. However, the SiO masers of the source at Q-band were too weak to be detected during our observations (due to their variable nature (\citealt{brand2020}). The recording rate was 1\,Gbps (2-bit sampling) where a total bandwidth of 256\,MHz was divided into eight 32-MHz intermediate frequency (IF) bands. Only left-hand circular polarization was received. All the data were correlated at the Daejeon hardware correlator installed in Korea Astronomy and Space Science Institute (KASI). 

In figure~\ref{fig:uv}(a) we display the \emph{uv}-coverage of M\,87 for Session-K. The baselines of KAVA+TMRT and KAVA+NSRT are shown with blue and red colors, respectively. The addition of TMRT to KaVA improves the \emph{uv}-coverage within 180\,M$\lambda$, while NSRT significantly extends the southeast-northwest baseline coverage by a factor of 2. Figure~\ref{fig:uv}(b) and (c) show the corresponding beam patterns with a naturally-weighted scheme for KaVA and KaVA+TMRT for Session-Q. Similarly, figure~\ref{fig:uv}(d)-(f) show the beam patterns for Session-K. In table~\ref{tab:imageP}-column(MSLL), we list some representative levels of maximum side lobes obtained from the beam patterns for each source, which are a good proxy to foresee the improvement of imaging performance. One can clearly see that the side lobe levels are significantly reduced by adding TMRT(+NSRT).

\subsection{Data reduction processes}
The EAVN data were calibrated in the standard manner of VLBI data reduction procedures  and under the guideline of EAVN data reduction. We used the National Radio Astronomy Observatory (NRAO) Astronomical Image Processing System (AIPS; \citealt{greisen2003}) software package for the initial calibration of visibility amplitude, bandpass and phase. As a large dish, the visibility amplitude of TMRT shows some offsets compared with that of KaVA. The system temperature information of NSRT was absent due to the malfunction of the data acquisition system. Subsequent imaging/CLEAN and self-calibration which were properly performed with the DIFMAP software (\citealt{shepherd1994}) can successfully recover the amplitude information from TMRT and NSRT. More details are described in the EAVN memo \footnote{https://radio.kasi.re.kr/eavn/data$\_$reduction.php}.

In the phase calibration process, solution intervals of 1\,min and 30\,s (with $SNR_{\rm ij}$ threshold 5) were used for Session-K and Session-Q, respectively, which are typical coherence time at each frequency. For comparison, in figure~\ref{fig:snr} we show baseline $SNR_{\rm ij}$ obtained by {\tt FRING} (on 3 baselines) for the point source 1219+044, for which the correlated flux densities are similar over the whole baselines. The measured SNR of KTN-TMRT at 22\,GHz were $\sim$4.4/$\sim$3.4 times higher than those of KTN-MIZ/KTN-KUS, which are consistent with our expectation under slightly rainy weather conditions at TMRT. At 43\,GHz, on the other hand, the measured SNR on KTN-TMRT was $\sim$4.2 times higher than KTN-MIZ. At 22\,GHz, the SNR of KTN-NSRT is comparable with that of intra-KaVA baselines. This is broadly consistent with our expectation given the typical sensitivity of NSRT. For M\,84, the fringes at 43\,GHz were not detected even with TMRT. This would be reasonable given the low core flux ($\sim$100\,mJy), relatively shorter integration time than Session-K. In figure~\ref{fig:amp17077K} and \ref{fig:amp17086Q}, we show the fully visibility amplitude of 1219+044, 3C\,273, M\,84 and M\,87 at 22/43\,GHz as a function of \emph{uv}-distance.

\begin{figure*}[htbp]
 \begin{center}
 \includegraphics[width=0.95\textwidth]{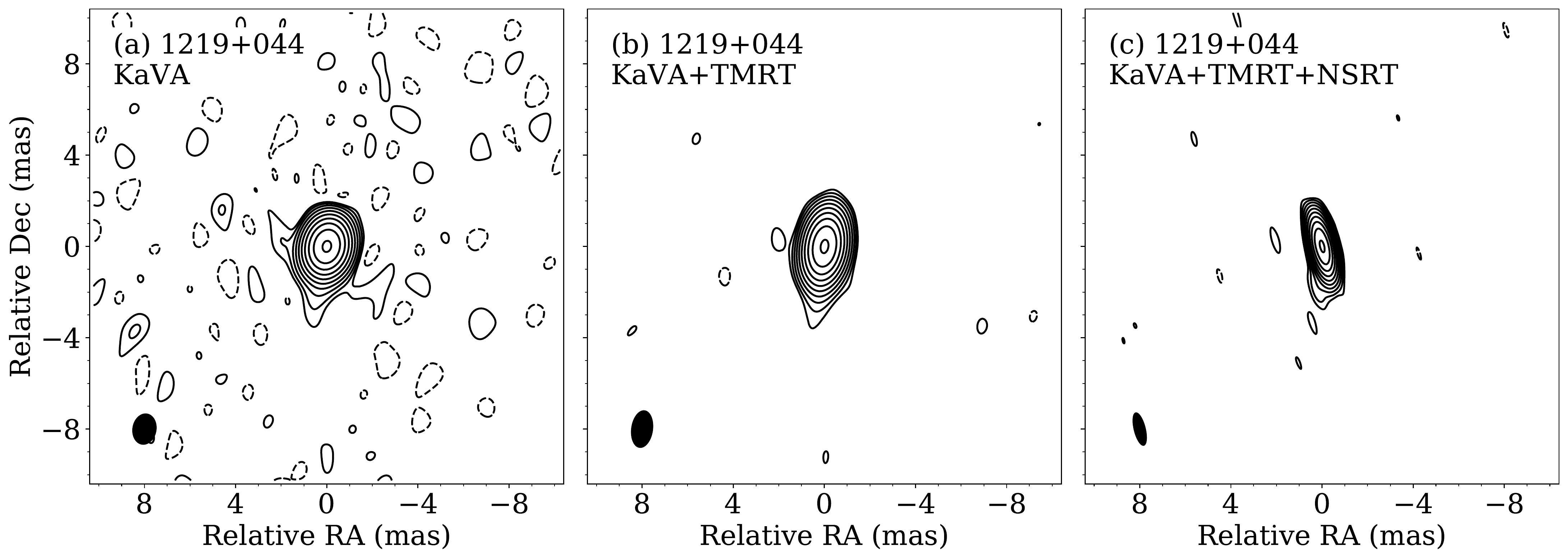}\\
  \includegraphics[width=0.95\textwidth]{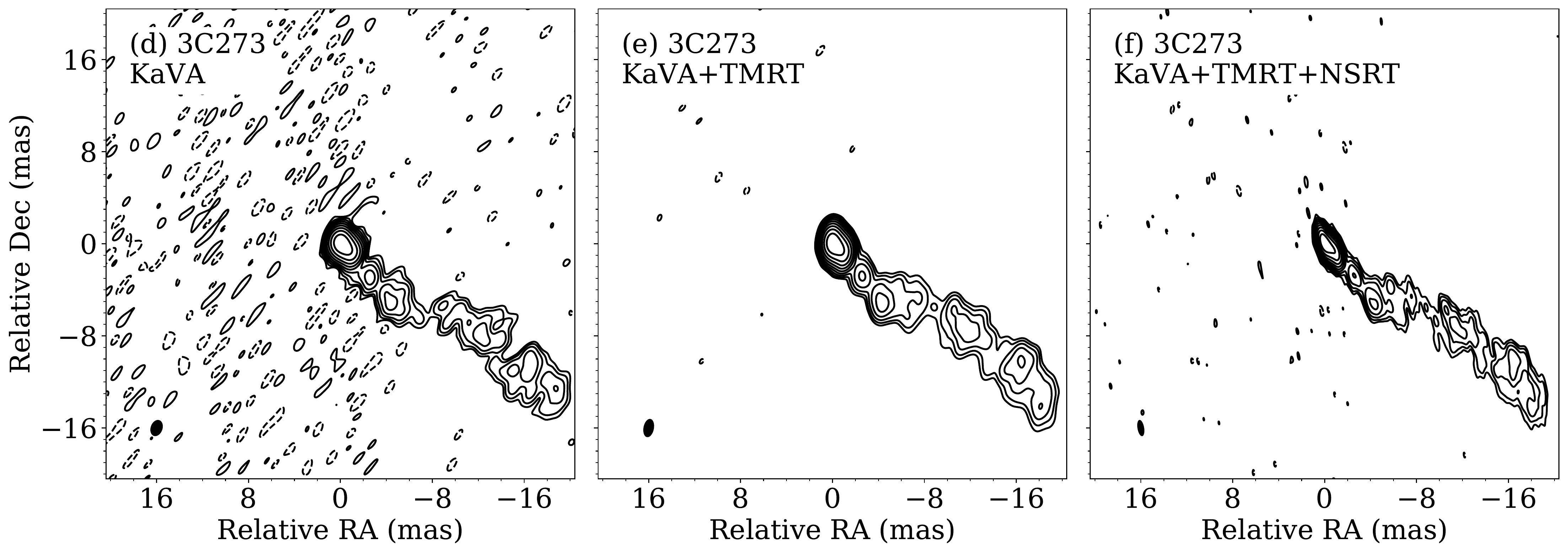}\
    \includegraphics[width=0.95\textwidth]{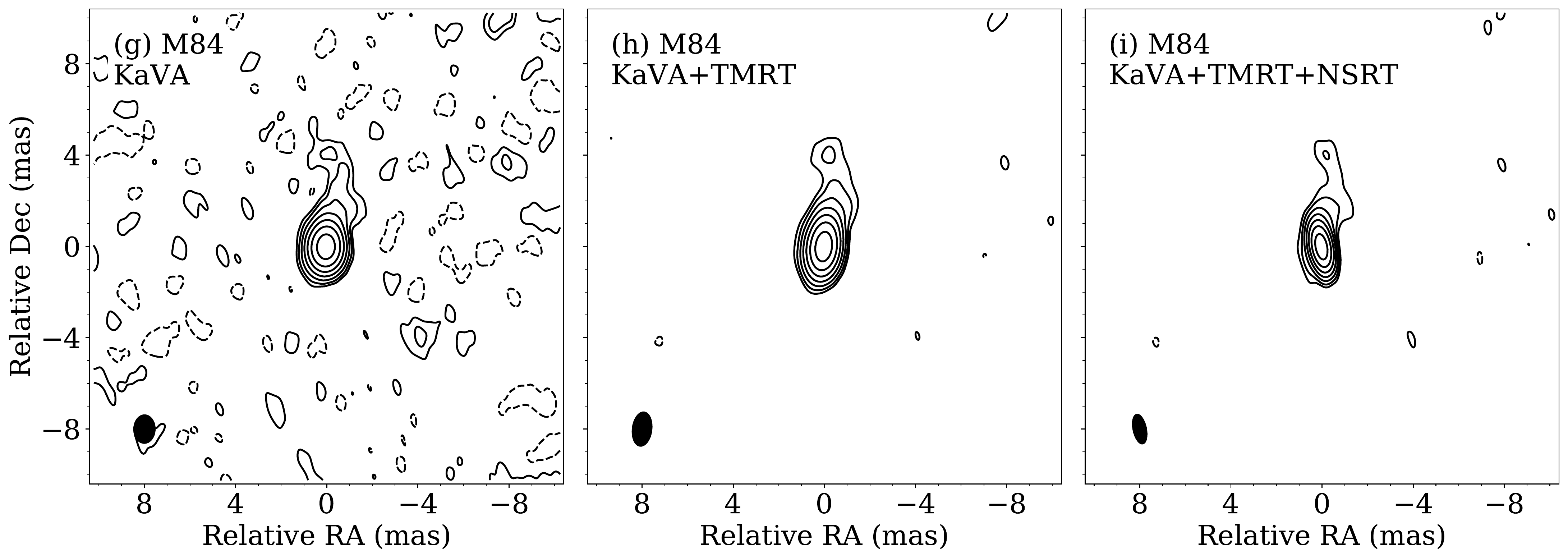}\\
  \includegraphics[width=0.95\textwidth]{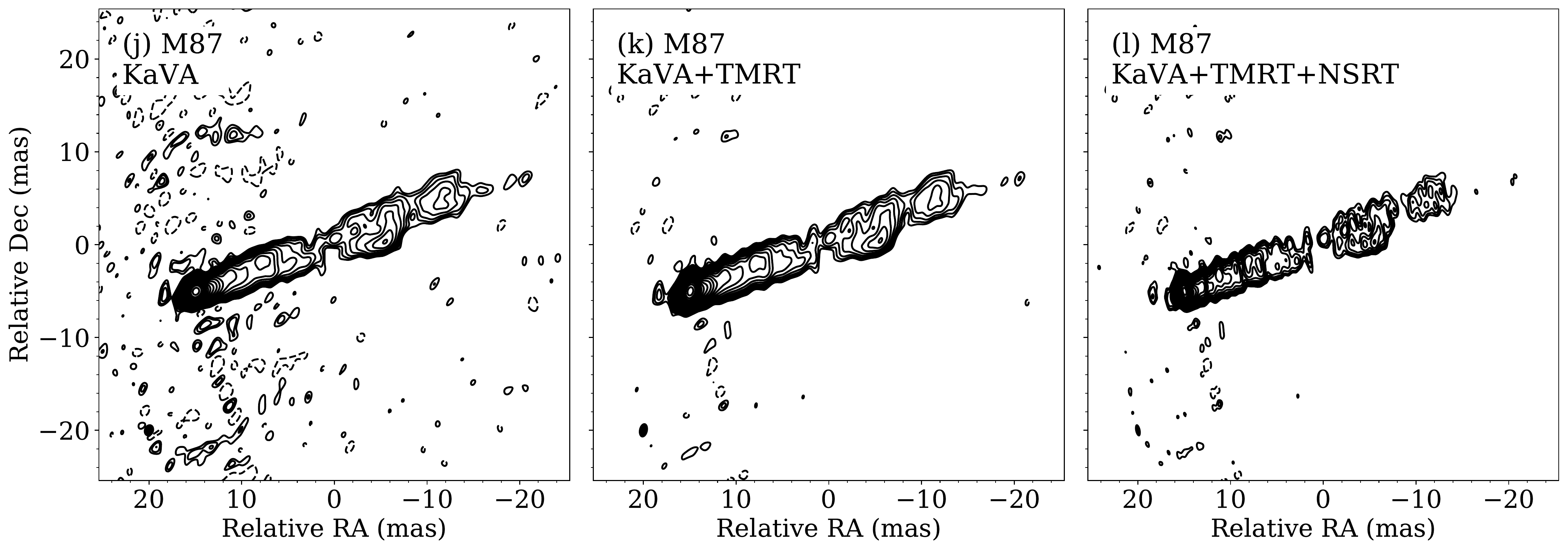}\\
 \end{center}
 \caption{Structure images of 1219+044 (a-c), 3C\,273 (d-f), M\,87 (g-i) and M\,84 (j-l): KaVA ({\it left}), KaVA+TMRT ({\it middle}) and KaVA+TMRT+NSRT ({\it right}) in Session-K. The first contour is consistent in both images for the same source, namely 1.02\,mJy/beam for 1219+044, 9.32\,mJy/beam for 3C\,273, 1.20\,mJy/beam for M\,87, 1.37\,mJy/beam for M\,84. The increase steps are (-1, 0, 1, 1.4, 2, 2.8...). A synthesized beam is indicated in the bottom-left corner of each panel.}
 \label{fig:compK}
\end{figure*}

\begin{figure}[ttt]
 \begin{center}
 \includegraphics[width=0.7\textwidth]{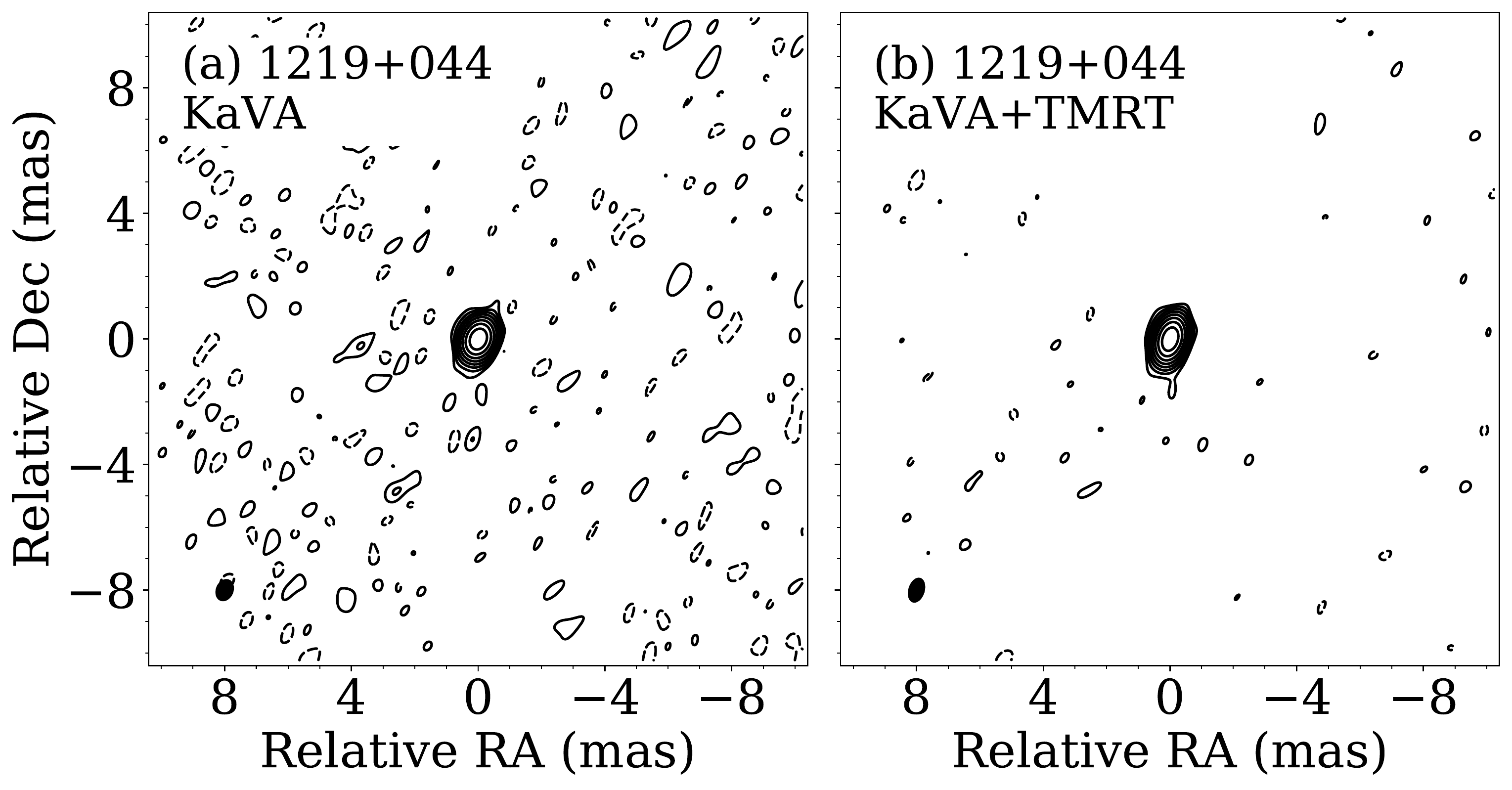}\\
  \includegraphics[width=0.7\textwidth]{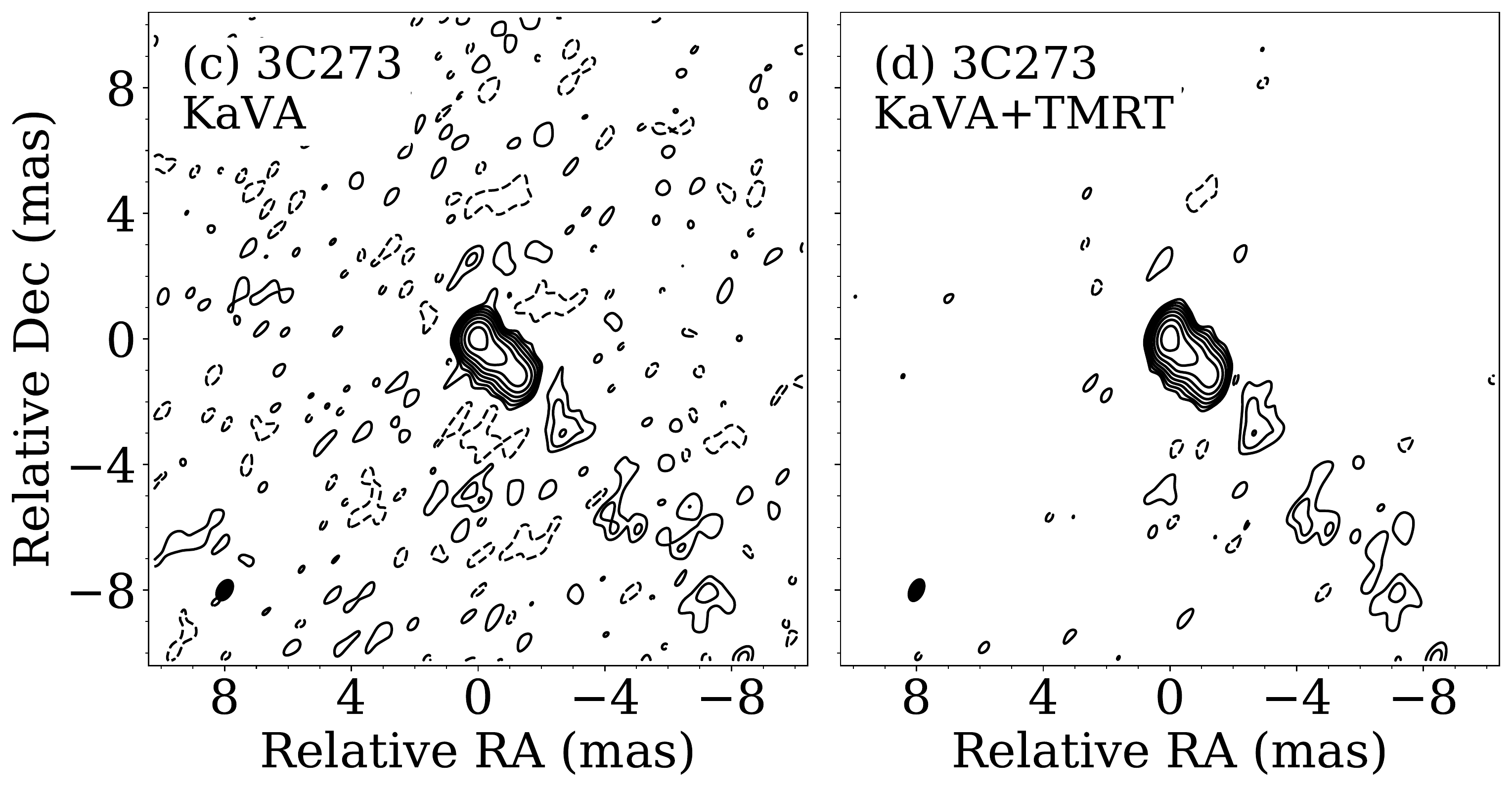}\\
  \includegraphics[width=0.7\textwidth]{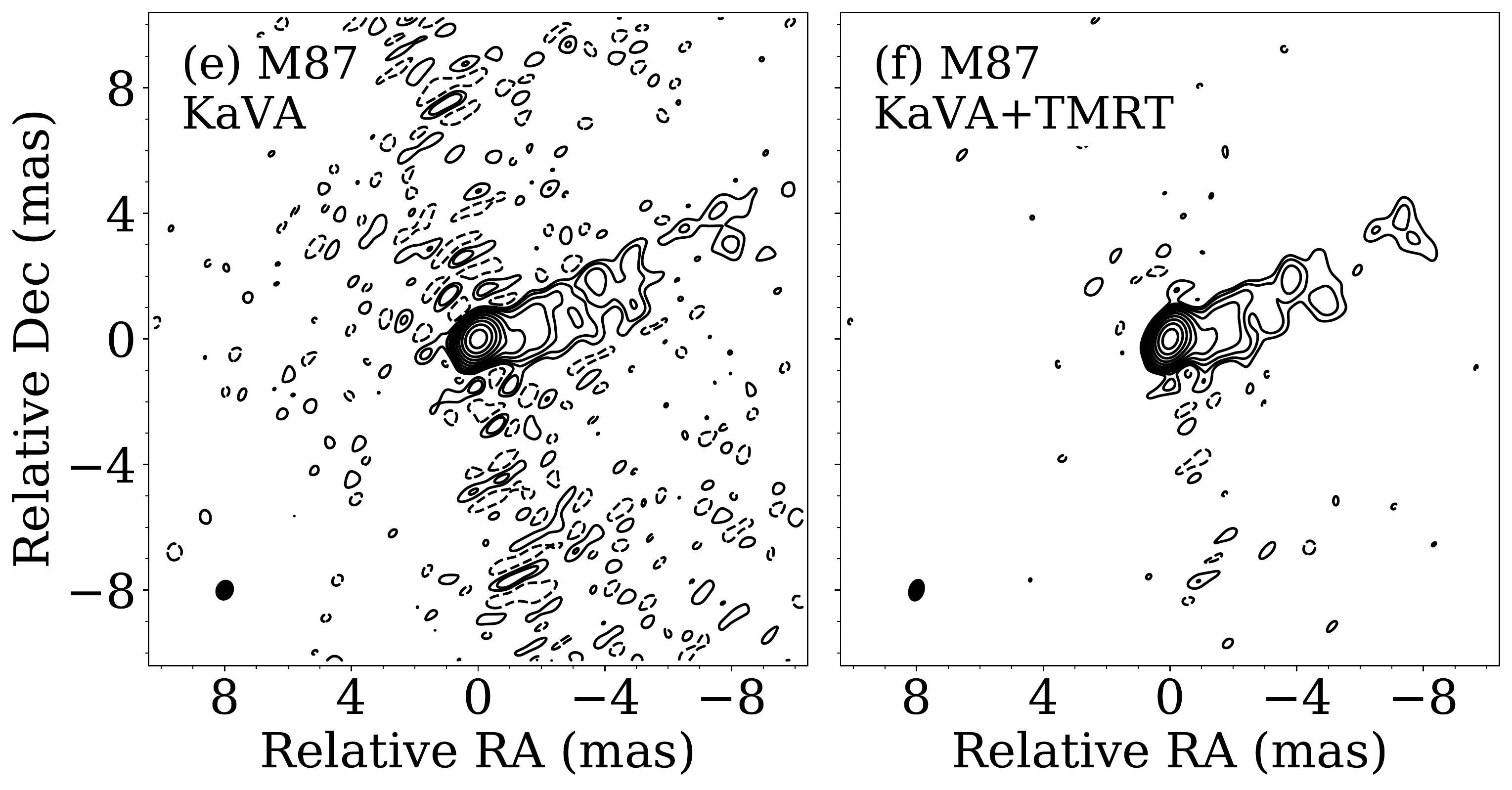}
 \end{center}
 \caption{Structure images of 1219+044 (a, b), 3C\,273 (c, d) and M\,87 (e, f): KaVA (a, c, e) and KaVA+TMRT (b, d, f) in Session-Q. The first contour is consistent in all images for the same source, namely 3.04\,mJy/beam for 1219+044, 20.19\,mJy/beam for 3C\,273, 1.49\,mJy/beam for M\,87. The increase steps are (-1, 0, 1, 1.4, 2, 2.8...). A synthesized beam is indicated in the bottom-left corner of each panel.}
 \label{fig:compQ}
\end{figure}

\begin{table}
\begin{center}
\caption[]{Image parameters of figure~\ref{fig:compK} and~\ref{fig:compQ} with natural weighting. $^a$ Total integration time on each source. $^b$ Image total flux density. $^c$ Array symbols same with table~\ref{tab:ims}. $^d$ Image rms noise level. Core (or off-core) region indicates the region within 5-mas (or over 40-mas) distance from the core. $^e$ Image peak intensity. $^f$ Dynamic range defined by $DR = I_{\rm peak}$/$I_{\rm{rms}}$. $^g$ The maximum side lobe levels are measured within the fringe space derived by the shortest baseline length, namely $\pm$5\,mas from the center for Session-K and $\pm$2.5\,mas from the center for Session-Q, along PA = +28/-152 $^\circ$.}\label{tab:imageP}
\setlength{\tabcolsep}{1.2pt}
\small
    \begin{tabular}{lcccclccccccl}
    \hline
    \hline
    Epoch  &Source&t$^a$& $I_{\rm{total}}^b$  & Array$^c$ & Beam size& \multicolumn{3}{c}{$I_{\rm{rms}}^d$ (mJy/beam)} &  $I_{\rm{peak}}^e$  & \multicolumn{2}{c}{DR$^f$}    &MSLL$^g$\\
           &      &(min) &  (mJy)&      & (mas$\times$mas, deg)  & Theo. & Core & Off-core    & (mJy/beam)    & Core & Off-core&\\
    \hline
    a17077a  &1219+044& 24 &607  & A1   & 1.38, 1.06, $-$10 & 0.76 &0.63  &0.65&  598  &  944  & 917 & 0.43\\
    (22\,GHz)&        &    &     & A2   & 1.66, 0.96, $-$8  & 0.42 &0.34  &0.37&  598  &  1759 & 1608& $-$0.05\\
             &        &    &     & A3   & 1.51, 0.53, 14  & 0.37 &0.34  &0.38&  594  &  1732 & 1561& $-$0.03\\
             &3C\,273 &36  &13067& A1   & 1.45, 1.03, $-$20 & 0.97 &4.53  &4.29&  5730 &  1264  & 1335&  0.41\\
             &        &    &     & A2   & 1.61, 0.91, $-$11 & 0.67 &3.11  &2.98&  5540 &  1783 & 1861& $-$0.03\\
             &        &    &     & A3   & 1.43, 0.58, 10  & 0.63 &3.37  &2.77&  5090 &  1511 & 1840& $-$0.01\\
             & M\,84  &  24&172  & A1   & 1.29, 0.98, $-$1  & 1.13 &0.93  &1.02&  148  &  159  & 145&  0.39\\
             &        &    &     & A2   & 1.55, 0.90, $-$7  & 0.59 &0.46  &0.48&  149  &  325  & 307& $-$0.03\\
             &        &    &     & A3 & 1.37, 0.63, 12  & 0.57 &0.45  &0.46&  146  &  326  & 316& $-$0.03\\
             & M\,87  &282 &2009 & A1 & 1.34, 1.08, $-$21 & 0.26 &0.67  &0.48&  1326 &  1980 & 2754& 0.36 \\
             &        &    &     & A2   & 1.56, 0.96, $-$12 & 0.16 &0.40  &0.29&  1313 &  3284 & 4544& $-$0.03\\
             &        &    &     & A3   & 1.35, 0.58, 12  & 0.15 &0.35  &0.25&  1154 &  3321 & 4586& $-$0.01\\
    \hline
    a17086a  &1219+044&  8 &692  & A1   & 0.73, 0.56, $-$22  & 1.84 &1.57  & 1.50  & 672  & 428 & 448 &  0.27 \\
    (43\,GHz)&        &    &     & A2   & 0.82, 0.52, $-$17  & 1.35 &1.01  & 0.95  & 672  & 662 & 707 &  $$-$$0.03\\
             & 3C\,273& 24 &8617 & A1   & 0.78, 0.52, $-$31  & 1.65 &9.04  & 8.11  & 3840 & 425 & 474 & 0.38\\
             &        &    &     & A2   & 0.81, 0.50, $-$25  & 1.56 &6.73  & 6.46  & 3840 & 571 & 634 & 0.13\\
             & M\,87  & 216&1577 & A1   & 0.68, 0.57, $-$23  & 0.37 &0.74  & 0.46  & 1180 & 1596 & 2565& 0.28\\
             &        &    &     & A2   & 0.74, 0.51, $-$17  & 0.31 &0.51  & 0.31  & 1180 & 2314 & 3764 &  0.01\\
  \noalign{\smallskip}\hline
\end{tabular}
\end{center}
\end{table}

\section{Results and discussion}
\label{sec:result}
\subsection{Structure images}

\begin{figure}[htbp]
 \begin{center}
  \includegraphics[width=0.7\textwidth]{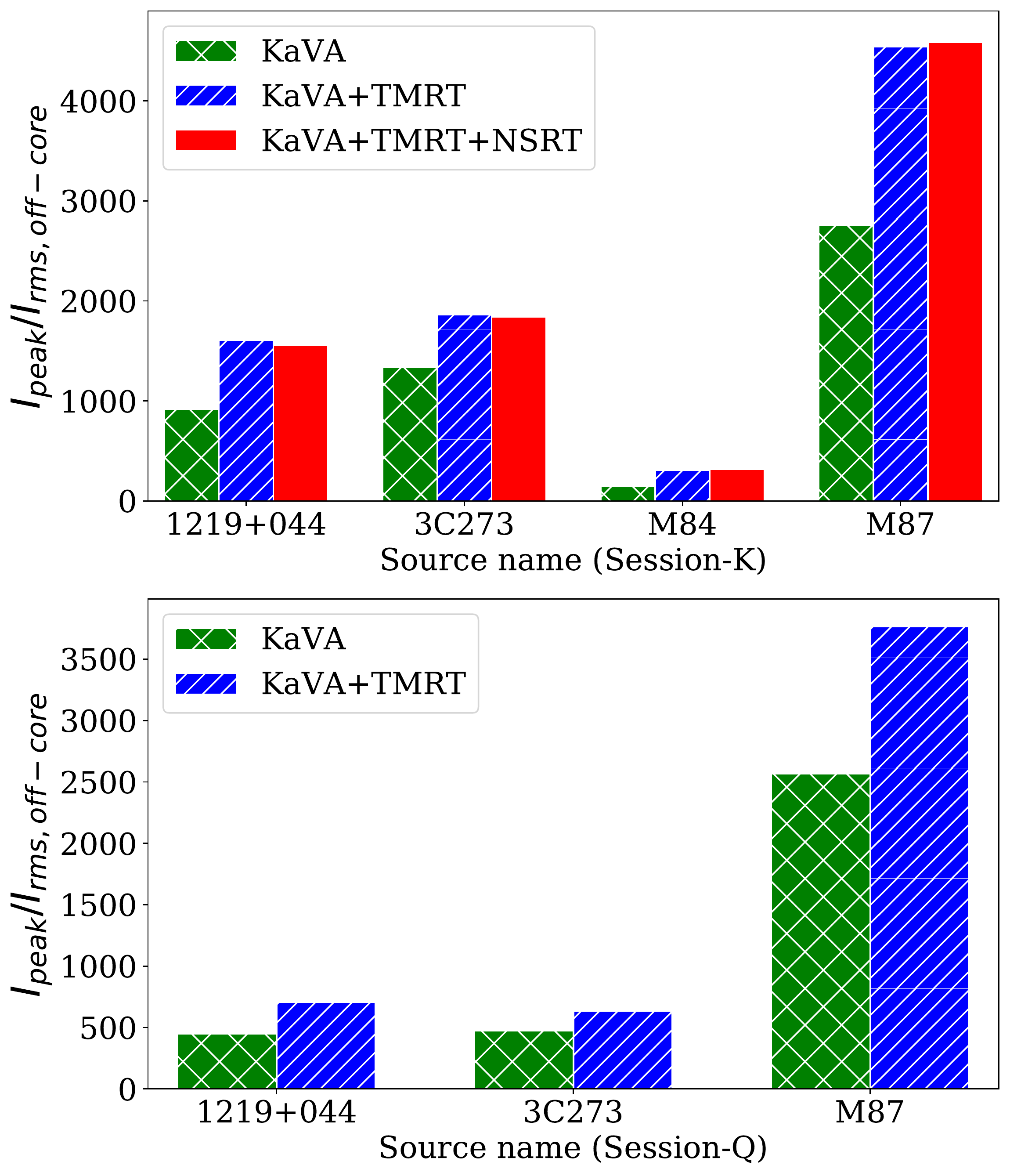}
 \end{center}
 \caption{Dynamic range measured in the off-core region for each source in Session-K (upper) and Session-Q (bottom). The detailed value are listed in table~\ref{tab:ratio}}
 \label{fig:DR}
\end{figure}

In figure~\ref{fig:compK} and \ref{fig:compQ}, we show the structure images of each source obtained from the Session-K and Session-Q, respectively. For each source, images obtained from KaVA, KaVA+TMRT, and KaVA+TMRT+NSRT are shown separately (with identical contour levels). Overall, the images indicate a significant improvement of image quality when TMRT and NSRT were added to KaVA, and the side lobes seen in KaVA images were greatly reduced for all the sources at both frequencies. The detailed image parameters and resulting image dynamic ranges ($DR = I_{\rm{peak}}$/$I_{\rm{rms}}$, where $I_{\rm{peak}}$ is the peak flux density and $I_{\rm{rms}}$ is image rms noise level.) are summarized in table~\ref{tab:imageP}. Here we calculated DR for two distinct regions. One is the core region which indicates the area within 5-mas distance to the core where the deconvolution errors usually dominate. The other one is the off-core region which represents the place over 40-mas distance to the core where the thermal noises dominate. The corresponding off-core $DR$ is also visualized in figure~\ref{fig:DR}. For all cases with different arrays and frequencies, the highest $DR$ were obtained for M\,87. This is simply because the total integration time of M\,87 was the longest and the corresponding \emph{uv}-coverage was much better than for the other sources.

For a proper comparison of relative improvement of image $DR$ among different arrays, a good measure would be the ratio of $DR$ rather than the absolute $DR$ for each image. In table~\ref{tab:ratio} we list the ratios of $DR$ which are calculated as follows: 

\begin{equation}\label{R21}
R_{21}=\frac{DR_{\rm A2}}{DR_{\rm A1}},
\end{equation} 
\begin{equation}\label{R31}
R_{31}=\frac{DR_{\rm A3}}{DR_{\rm A1}},
\end{equation} where A1, A2, and A3 indicate the array with KaVA, KaVA+TMRT and KaVA+TMRT+NSRT, respectively. Overall, $DR$ is increased significantly when TMRT is added to KaVA, with actual improvement factors varying from $\sim$34\% to $\sim$112\%. The highest relative improvement was achieved in M\,84, which is the weakest source in our sample. Then the point-like source 1219+044 comes second followed by M\,87. The improvement was the lowest in 3C\,273. The east-west angular resolution of the images including NSRT is twice better than KaVA+TMRT images. However, the $DR$ does not become worse when the resolution becomes better.  

\subsection{Point-like source 1219+044}
\label{sec:result1219}

The EAVN images of 1219+044 shown in figure~\ref{fig:compK}(a)-(c) and \ref{fig:compQ}(a)-(b) are characterized by a quasi-unresolved structure at both frequencies, which is consistent with the flat visibility amplitude distributions in the \emph{uv} domain (figure~\ref{fig:amp17077K} and \ref{fig:amp17086Q}). For all images from different arrays, we confirmed that the resulting image rms levels were close to thermal limits, as expected for such a simple source structure.

The improvement of image $DR$ for this source is remarkable: by adding TMRT(+NSRT) to KaVA, a factor of $\sim$1.7--1.9 and $\sim$1.5--1.6 enhancement was seen at 22\,GHz and 43\,GHz, respectively. Thanks to the significant improvement of image quality, the KaVA+TMRT(+NSRT) image(s) at 22\,GHz allows us to detect a hint of weak extended emission towards the south, which is consistent with the jet emission seen in the literature VLBI images (\citealt{lister2019}). 

\begin{table}
\begin{center}
\caption[]{ Improvement factors of image $DR$ compared to KaVA. Calculation equations are equations~\ref{R21} and \ref{R31}.}\label{tab:ratio}


 \begin{tabular}{llcccccc}
  \hline\noalign{\smallskip}
      Epoch &Source & \multicolumn{2}{c}{Core region} &\multicolumn{2}{c}{Off-core region} \\
       & &$R_{21}$  &  $R_{31}$    &$R_{21}$  &  $R_{31}$  \\
    \noalign{\smallskip}\hline
    a17077a  &1219+044  &1.86 &1.83 & 1.75& 1.70\\
    (22 GHz)    &3C\,273    &1.41 &1.20  &1.39 &1.38 \\
        &M\,84      & 2.05& 2.05 &2.12 &2.18 \\
        &M\,87      & 1.66&1.68  &1.65 &1.66\\
    \noalign{\smallskip}\hline
    a17086a  &1219+044  & 1.55&--& 1.58 &--\\
    (43 GHz)    &3C\,273    & 1.34& --& 1.34 &--\\
        &M\,87      & 1.45&-- & 1.47 &--\\
  \noalign{\smallskip}\hline
\end{tabular}
\end{center}
\end{table}

\subsection{Kink of 3C\,273 jet}

As shown in figures~\ref{fig:compK}(d)-(f) and~\ref{fig:compQ}(c)-(d), the addition of TMRT to KaVA increased the image dynamic range of 3C\,273 by $\sim$40\% at both frequencies. Since the source structure of 3C\,273 is complex with a bright core, the resulting image rms noise levels were $\sim$3 times larger than the thermal noise limit, indicating that EAVN imaging of 3C\,273 is highly dynamic range limited.  

Another challenge of 3C\,273 imaging is that, due to the complicated source structure, the calibrated visibility data could leave larger systematic errors than the case for a simple structure source. Hence, the image $DR$ improvement of the KaVA+TMRT+NSRT at 22\,GHz was relatively modest. Nevertheless, a factor of 2 enhancement of the east-west resolution allowed us to resolve the fine-scale structure of the core and jet of 3C\,273.

The jet in quasar 3C\,273 holds a well-known appealing structure which appears to be bent at mas scales (\citealt{readhead1979})while is highly collimated at arcsecond scales (\citealt{bahcall1995}). Previous high-resolution VLBI images at $\leq$22\,GHz consistently indicate a characteristic persistent ``kink'' morphology at $\sim$7--10\,mas from the core, where the bright edge of the jet flips from one side to the other (\citealt{zensus1988, bruni2017, akiyama2018, lister2019, zensus2020}). Thanks to the image $DR$ improvement by a factor of $\sim$1.4, the addition of TMRT (and NSRT) to KaVA is better tracing the bending structure of 3C\,273 in 7--10\,mas regions (see figures~\ref{fig:compK}(d)-(f)).

\subsection{Weak jet of M\,84}

The addition of TMRT/NSRT to KaVA allowed us to robustly detect weak signals of this source at 22\,GHz. The EAVN images clearly ($>$10$\sigma$) detected a weak jet structure towards the north thanks to the factor of 2 improvement of image $DR$ compared to KaVA. In the KaVA+TMRT+NSRT image, there is a possible hint of slight extension also towards the south, which could be associated with the counter-jet as seen in the literature VLBI images (\citealt{giovannini2001, ly2004}). A deeper imaging observation with a longer integration time may confirm this structure in future.

\subsection{Counter jet feature of M\,87}

\begin{figure}[htbp]
 \begin{center}
  \includegraphics[width=0.8\textwidth]{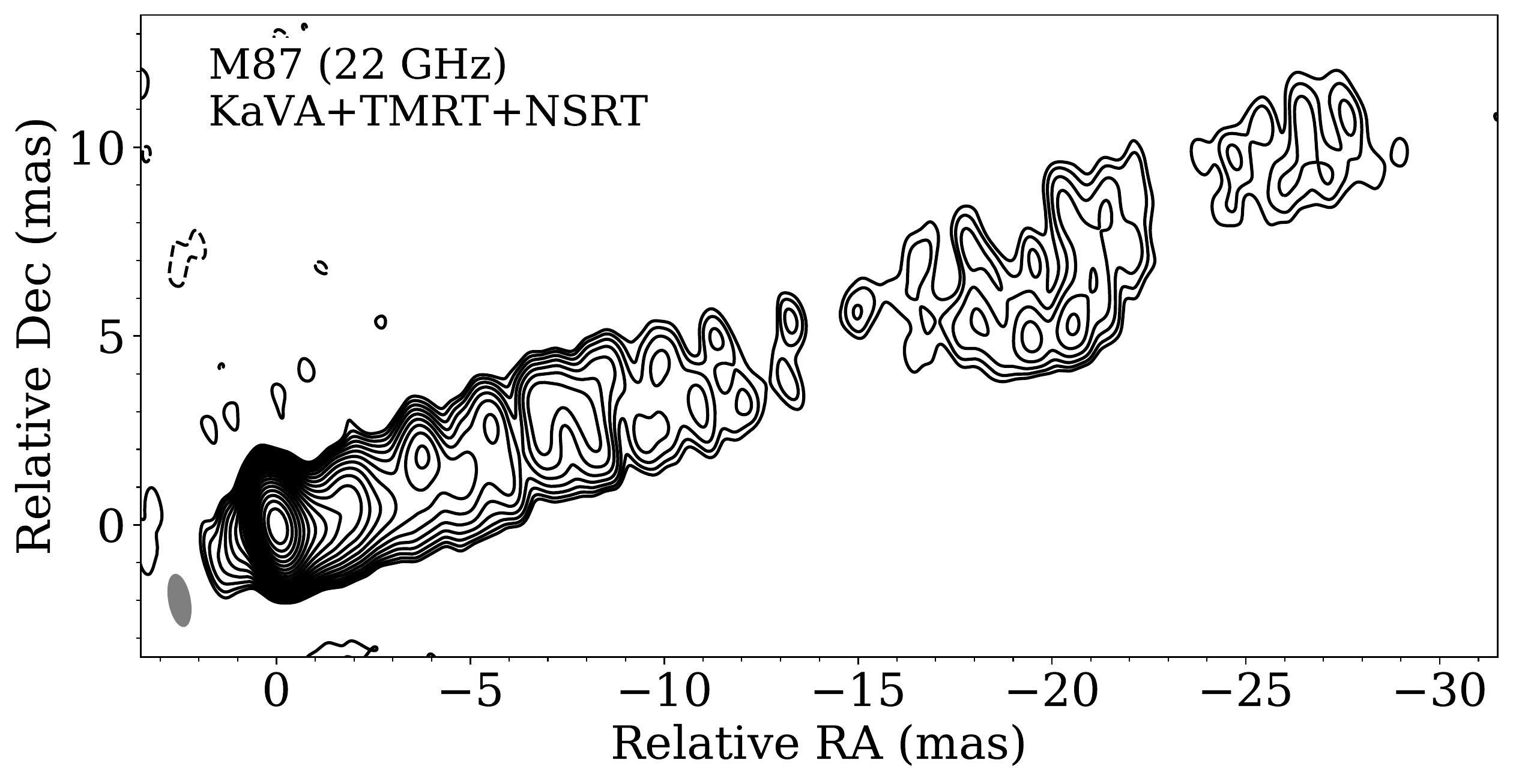}
    \end{center}
 \caption{KaVA+TMRT+NSRT image of M\,87 in Session-K covering the whole jet up to $\sim$30\,mas from the core. Contour levels are scaled as (-1, 0, 1, 1.4, 2, 2.8...)$\times$1.20 mJy/beam. The synthesized beam is shown in the bottom-left corner of the figure.}
 \label{fig:contK}
\end{figure}

For M\,87, the resulting dynamic range of EAVN images reached $>$4500 at 22\,GHz (figure~\ref{fig:compK}(l)) and $>$3700 at 43\,GHz (figure~\ref{fig:compQ}(f)), thanks to the longest on-source time among the 4 sources. As a result, the extended jet emission was clearly detected up to 30\,mas (figure~\ref{fig:contK}) at 22\,GHz, which is important to investigate the collimation profile over long distances. The $DR$ improvement of EAVN images with respect to KaVA is more significant at 22\,GHz ($\sim$70\%) than at 43\,GHz ($\sim$50\%). This should be mainly due to the overall higher array sensitivity as well as the longer on-source time at 22\,GHz. 

The significant enhancement of sensitivity and east-west angular resolution by adding TMRT+NSRT enabled us to robustly detect/resolve the ``counter-jet" at 22\,GHz, which is extending up to $\sim$2\,mas at the eastern side of the core (see figure~\ref{fig:compK}-l). The detection of this feature was cross-checked by different data analysis to make sure that its appearance is not an artifact. The counter-jet of M\,87 was firstly suggested in \citet{ly2004} and later confirmed in a number of VLBA images at various frequencies (\citealt{ly2007, kovalev2007, hada2016, kim2018, walker2018}). Imaging of the counter-jet near the core is vital to better understand the properties of jet-launching regions (\citealt{hada2016, kim2018, walker2018}). In particular, accurate measurements of the jet-to-counter-jet brightness ratio (BR) allows us to constrain some key parameters such as the viewing angle and launching velocity, based on the standard theory of Doppler-boosting/deboosting of relativistically moving plasma.  

At 43\,GHz, thanks to the higher angular resolution transverse to the jet axis, the KaVA+TMRT image clearly resolved the limb-brightening structure that is well known in this jet. Interestingly, the southern limb of the jet appears to be brighter than the northern limb, which might be connected to the asymmetric brightness distribution the black-hole shadow seen in the EHT image (\citealt{eht2019a}). 

\section{EAVN data status after 2017}
\label{sec:2021}
While here we presented the EAVN data obtained from the commissioning observations in 2017, the EAVN array has been continuously evolving since then. From 2018, EAVN officially started open-use observations with KaVA+TMRT+NSRT being a core array, and now EAVN is in its regular operation \footnote{https://radio.kasi.re.kr/eavn/about$\_$eavn.php}. Our results of array performance evaluation presented here (based on 2017 data) should serve as a useful guidance/reference for the calibration of any EAVN data obtained after 2017.  In fact, we analyzed some of more recent KaVA+TMRT+NSRT data obtained by an open-sky program, and we confirmed good consistency of the array performance as reported here. Various scientific results based on EAVN data are currently being produced (e.g., gamma-ray bursts, \citealt{an2020}). The detailed results of M\,87 and SgrA obtained from recent EAVN observations will be reported in Cui et al. 2021 (in preparation) and Cho et al. 2021 (in preparation). 

Concurrently, to further expand the baseline length and imaging capability, a joint effort of VLBI operation between EAVN and and radio telescopes in Italy is actively ongoing. This forms the East-Asia-To-Italy-Nearly-Global (EATING) VLBI array, which for the first time realizes a global VLBI array operating on a regular basis (\citealt{hada2019}). The EATING VLBI array is very powerful to intensively monitor the kinematics of the innermost regions of AGN jets. Moreover, joint observations with other European (e.g., Yebes), Russian and Australian are also under commissioning. All of these efforts on global scales are actively ongoing and more scientific outcomes with these efforts will be yielded in the coming years.

\section{Summary}
\label{sec:summary}
In this paper, we reported the first imaging results of bright AGN jets with KaVA+TMRT(+NSRT) that serves as a core array of EAVN. Compared with only KaVA, we confirmed that the image dynamic range for a point source (1219+044) can be improved by $>80\%$ and $>50\%$ at 22 and 43\,GHz, respectively. For the source with extended jets M\,87, KaVA+TMRT+NSRT successfully recovered their rich structures at image $DR$ over 4500, thanks to the significant enhancement of sensitivity by TMRT and E-W angular resolution by NSRT. This demonstrates the excellent capability of EAVN for study of jet formation and collimation of powerful relativistic jets. We also imaged a relatively weak ($\sim$100\,mJy) source with short integration time. 

In the next few years and beyond, we will see several new stations in East Asia and Southeast Asia, such as extended-KVN (eKVN; e.g., \citealt{jung2015}), QiTai Telescope (QTT; e.g., \citealt{xu2016}) and Thai National Radio Telescope (TNRT; e.g., \citealt{jar2017}), which are all capable of 22/43\,GHz. In the meantime, plans of EAVN experiments at low frequencies (1--10\,GHz) together with the FAST 500-m telescope (e.g., \citealt{nan2006}) are being actively considered. Ultimately, the growing EAVN can be connected to other cm-VLBI networks such as EVN, LBA and VLBA. This will facilitate the realization of a truly global VLBI array at centimeter wavelengths. 

\begin{acknowledgements}
We thank the anonymous referee for her/his careful review and suggestions that improved the manuscript. We acknowledge all staff members and students at TMRT, NSRT, KVN and VERA who supported the operation of the array and the correlation of the data. TMRT is operated by Shanghai Astronomical Observatory. NSRT is operated by Xinjiang Astronomical Observatory. KVN is a facility operated at by the Korea Astronomy and Space Science Institute. VERA is a facility operated at National Astronomical Observatory of Japan in collaboration with associated universities in Japan. This work is supported by The Graduate University for Advanced Studies (SOKENDAI). Y.C. is supported by the Japanese Government (MEXT) Scholarship. This work is supported by JSPS KAKENHI Grant Numbers JP18K03656 (M.K.), JP18H03721 (K.N., K.H. and M.K.), JP19H01943 (K.H., F.T. and Y.H.) and JP18KK0090 (K.H. and F.T.). K.H. is supported by the Mitsubishi Foundation (grant number 201911019). J.P. is supported by an EACOA Fellowship awarded by the East Asia Core Observatories Association, which consists of the Academia Sinica Institute of Astronomy and Astrophysics, the National Astronomical Observatory of Japan, the Center for Astronomical Mega-Science, the Chinese Academy of Sciences, and the Korea Astronomy and Space Science Institute. J.P. and I.C. acknowledge the financial support from the National Research Foundation (NRF) of Korea via Global Ph.D. Fellowship Grant 2014H1A2A1018695 and 2015H1A2A1033752, respectively. S.T. acknowledges support from the NRF via Grant 2019R1F1A1059721. This work is supported by the Major Program of the National Natural Science Foundation of China (NSFC, Grant No. 11590780, 11590784), the Knowledge Innovation Program of the Chinese Academy of Sciences (Grant No. KJCX1-YW-18), the Scientific Program of Shanghai Municipality (08DZ1160100), and Key Laboratory for Radio Astronomy, CAS. W.J. acknowledges support from NSFC under Grant No. 11803071. L.C. is supported by the National Key R\&D Program of China under Grant No. 2018YFA0404602, the CAS 'Light of West China' Program under Grant No. 2018-XBQNXZ-B-021, the NSFC under Grant No. U2031212 and 61931002 and the Youth Innovation Promotion Association of the CAS under Grant No. 2017084. J.C.A acknowledges support from Fundamental Research Grant Scheme (FRGS) FRGS/1/2019/STG02/UM/02/6. R.-S. Lu is supported by the Max Planck Partner Group of the MPG and the CAS and acknowledges the support by the Key Program of the National Natural Science Foundation of China (NSFC grant No. 11933007) and the Research Program of Fundamental and Frontier Sciences, CAS (grant No. ZDBS-LY-SLH011).
\end{acknowledgements}

\bibliographystyle{raa}
\bibliography{mybib}
\label{lastpage}

\end{document}